\documentclass[12pt]{elsart}
\usepackage{color}
\usepackage{graphicx}
\def\be{\begin{eqnarray}}
\def\ee{\end{eqnarray}}
\def\bc{\begin{center}}
\def\ec{\end{center}}
\def\om{\omega}

\def\prt{\partial}

\def\lsim{\stackrel{\scriptstyle <}{\phantom{}_{\sim}}}
\def\gsim{\stackrel{\scriptstyle >}{\phantom{}_{\sim}}}

\def\rmd{{\rm d}}
\begin{document}
\begin{frontmatter}
\title{Equation of State for
 Hot and Dense Matter: $\sigma$-$\om$-$\rho $ Model
 with Scaled Hadron Masses and
Couplings\\
}
\author[JINR]{A.S.~Khvorostukhin},
\author[JINR,GSI]{V.D.~Toneev}
\and \author[MEPHI,GSI]{D.N. Voskresensky}
\address[JINR]{Joint Institute for Nuclear Research,
 141980 Dubna, Moscow Region, Russia}
\address[GSI]{GSI, Plankstra\ss{}e 1, D-64291 Darmstadt, Germany}
\address[MEPHI]{Moscow Engineering Physical Institute,\\ Kashirskoe
  Avenue 31, RU-115409 Moscow, Russia}

\begin{abstract}
  The proposed earlier relativistic mean-field model  with hadron masses and
coupling constants depending on the $\sigma$-meson field is
generalized  to finite temperatures. Within this approach we
simulate the in-medium behavior of the hadron masses motivated by
the Brown-Rho scaling.  The high-lying baryon resonances and boson
excitations as well as excitations of the $\sigma$, $\om$ and
$\rho$ fields interacting via mean fields are incorporated into
this scheme. Thermodynamic properties of hot and dense hadronic
matter are elaborated with the constructed equation of state. Even
at zero baryon density, effective masses of
$\sigma$-$\om$-$\rho$-$N$ excitations abruptly drop down for
$T\gsim 170$~MeV and reach zero at  a critical temperature
$T=T_{c\sigma}\sim 210$~MeV. Below $T_{c\sigma}$ (at $T\sim
190$~MeV) the specific heat gets a peak like  at crossover. We
demonstrate that our EoS can be matched with that computed on the
lattice for high temperatures provided  the baryon resonance
couplings with nucleon are partially suppressed. In this case the
quark liquid would masquerade as the hadron one. The model is
applied to the description of heavy ion collisions in a broad
collision energy range. It might be especially helpful
 for studying phase diagram in the region near possible phase transitions.

\end{abstract}\end{frontmatter}

\section{Introduction}

The investigation of thermodynamic properties and phase structure
of strongly interacting nuclear matter at high baryon densities
and  temperatures has recently become important in view
of plans to construct new accelerator facilities (FAIR) at GSI
Darmstadt for covering the (5-35) AGeV heavy-ion energy
range~\cite{SST}. Increasing interest to this energy region
is emphasized by the recent proposal for a low-energy campaign at
RHIC aimed at identification of the critical
end-point~\cite{RHIC_low} and by current discussions about the
feasibility for searching the quark-hadron mixed phase at the
Nuclotron-based collider  (JINR, Dubna)~\cite{NICA}.

Theoretical predictions for critical baryon densities and
temperatures of phase transitions  depend sensitively on the
equation of state (EoS)  of both the hadronic matter  and the
quark-gluon matter at high densities and temperatures in the
nonperturbative regime.  Here we focus on studying  the hadronic
EoS. An EoS of hadronic matter should satisfy experimental
information extracted from the description of global
characteristics of atomic nuclei such as the saturation density,
binding energy per particle,  compressibility,  asymmetry energy
and some other. Some constraints on hadronic models of EoS follow
from  analysis of elliptical flow {and  $K^+$ data}
in heavy ion collisions (HIC) . In addition to these constraints
astrophysical bounds on the high-density behavior of
$\beta$-equilibrium neutron star matter were applied in the recent
paper~\cite{Army}.

{Obviously relativistic effects are important under
extreme conditions of high densities and temperatures.
Microscopically based  approaches, as the
Dirac-Brueckner-Hartree-Fock method, see \cite{DBHF}, are very
promising but need rather involved calculations. Uncertainties are
getting larger with increase of the baryon density and
temperature. E.g., three-body forces are to large extend
unconstrained. Their poorly-known isospin and temperature
dependencies introduce additional sources of uncertainty. In the
quark-meson-coupling models the nuclear system is represented as a
collection of quark bags. The interactions are generated by the
exchange of $\sigma$-, $\omega$- and $\rho$-mesons treated on the
mean-field level  with quarks, for recent review see \cite{STT}.
These models might be especially useful for the study of the quark
deconfinement phase transition. But they involve uncertainties due
to phenomenological description of the nucleon structure. As a
more economical approach for describing global properties of
nuclear matter at the above extreme conditions the relativistic
mean field (RMF) approach is often used where baryons interact
with $\sigma$, $\omega$ and $\rho$ mean fields. Parameters of the
model are extracted from the comparison with the experimental
data. First RMF models used a minimal coupling of nucleons with
$\sigma$, $\omega$ and $\rho$ mesons \cite{Wal}. However it proved
to be insufficient to appropriately describe experimental data.
Therefore non-linear self-interactions of the $\sigma$ meson have
been introduced, see \cite{non-lin}. This approach was latter
extended to meson fields \cite{mes}. As an alternative, RMF models
with density dependent nucleon-meson couplings were developed
\cite{alt}. They allow  a more flexible description of the medium
dependence. Many other extensions of the models have been
considered, e.g.  models including $SU(3)$ symmetry \cite{B}. It
is not our goal here to study and compare different models with
each other. Most of the models were developed in order to describe
a specific domain of nuclear physics. Their validity in other
regions of nuclear physics either was not considered  or they
failed to describe them. In Ref.~\cite{Army} a general testing
scheme was  developed to apply the models  to all known nuclear
systems. It has been shown that among other phenomenological
models studied there, the RMF model of the EoS
 suggested in \cite{KV04}  proved
to be one of the most efficient model satisfying majority of the
existing constraints at zero temperature (see KVOR EoS in Table 5
in Ref.~\cite{Army}). Therefore focusing  on the further
application to HIC,  the given model will be generalized here to
finite temperatures. }

Following Ref. \cite{KV04} we assume a relevance of the  (partial)
chiral symmetry restoration  at high baryon densities and/or
temperatures ~\cite{chiral} manifesting in form of  the Brown-Rho
scaling hypothesis \cite{BR}:  Masses and coupling constants of
all hadrons decrease with a density increase  in approximately the
same way. {Note that
most of the models use the constant $\sigma$, $\om$, $\rho$
effective masses.} Some models introduce field interaction terms
leading to an increase of the $\sigma$, $\om$, $\rho$ effective
masses with the increasing nucleon density, e.g. \cite{HP,FST}.
Contrary, Ref.~\cite{KV04} follows the Brown-Rho scaling
hypothesis  and scales the quadratic (mass) terms of $\sigma$,
$\omega$ and $\rho$ fields as well as the baryon mass by a
universal scaling function $\Phi$. The scaling function $\Phi$ is
assumed to be  dependent on the $\sigma$ mean-field. This provides
thermodynamical consistency of the model. In order to obtain a
reasonable EoS, the meson-nucleon coupling constants should be
also scaled with the $\sigma$ mean field. Differences in the
scaling functions for the effective masses of $\omega$- and
$\rho$-fields and their couplings to a nucleon allow one to get an
appropriate density-dependent behavior of both the total energy
and the nuclear asymmetry energy, in agreement with constrains
obtained from  measurements of neutron star masses and surface
temperatures~\cite{Army}.

Our main goal  here is to construct some effective model of EoS
that would incorporate the decrease of hadron masses and couplings
with increase of the baryon density $n_B$ and temperature $T$
and, simultaneously, would fulfill various constraints
 known from analysis of atomic nuclei, neutron stars
 and HIC. Our consideration is based on the
generalization of  the KVOR model~\cite{Army} to finite
temperatures. We  aim to test its suitability to  description of
properties of hot and dense matter formed in HIC. Besides the
nucleon and meson mean fields we include low-lying non-strange and
strange baryon resonances ($N$, $\Delta$, $\Lambda$, $\Sigma$,
$\Xi$, $\Sigma^*$, $\Xi^*$,  and $\Omega)$, meson excitations
$\sigma (600)$, $ \rho (770)$,  $\omega (782)$ constructed on the
ground of mean fields , and the (quasi)Goldstone excitations $\pi
(138)$, $K(495)$, $\eta (547)$. We add here their high mass
partners in the $\rm{SU(3)}$ multiplet $K^* (892)$, $\eta^{'}
(958)$ and $\varphi (1020)$. All corresponding antiparticles are
also included. All states are treated within quasiparticle
approximation.

We restrict ourselves by taking into account only the large $N_c$
ground states of baryons and mesons~\cite{Mano98}. Generalization
to higher mass resonances is, of course, straightforward. However
we will drop them from our consideration being guided by the
following arguments: The next in mass state that does not enter
the multiplet is the $\Lambda (1405)$ state. It manifests in the
kaon scattering data as a quasimolecular state near the
kaon-nucleon threshold. However in description of a broad kaon
energy region under discussion in this work the $\Lambda (1405)$
hyperon does not manifest as a pole-like term \cite{KV03}. Ref.
\cite{WKW} presented arguments that $\Lambda (1405)$ should
dissolve in matter. Thereby  we will exclude $\Lambda (1405)$ from
our study. Moreover as it is conjectured  in Ref.~\cite{LK02},
higher mass baryon resonances can be understood as composite
particles. Their widths in matter are expected to be quite large.
Ref.~\cite{LK02} reproduces particle scattering data assuming that
the lowest baryon octet and decouplet are only relevant degrees of
freedom. Therefore we do not incorporate the higher resonances
within our quasiparticle model. Besides, the higher mass particles
are considered, the less they contribute to thermodynamics, and
the less one knows about their interactions.  Following above
argumentation and in order not to complicate consideration by
introducing dependencies on unknown parameters we accordingly cut
the baryon particle set.

Although   free $\sigma$-, $\rho$-, $\omega$-mesons are rather
heavy, their effective masses in matter essentially decrease.
Therefore we include excitations of these fields as well.
Sometimes the $\delta [a_0 (980)]$-meson is incorporated in the
RMF scheme~\cite{LG}. Its role in RMF models is similar to that of
the $\rho$-meson except that inclusion of $\delta [a_0
(980)]$-meson allows for mass splitting between proton and
neutron. Since $\delta [a_0 (980)]$ coupling constants to other
particles are unknown and the mass is larger than that for $\om$
we do not incorporate $\delta$ in our scheme.

To construct a practical description,  the particle interaction
with $\sigma$-, $\omega$- and $\rho$-meson fields is treated only
in the mean-field approximation. The fermion-fermion hole loop
diagrams for  boson propagators and the boson-fermion loop
diagrams for fermion propagators are disregarded. Thereby we omit
the  $p$-wave pion-baryon  and kaon-baryon interaction effects,
though these effects are important in high-baryon density
regime~\cite{KV03,MSTV90,KVK96}. At high temperatures the
fermion-antifermion loops in boson propagation and fermion-boson
loops in boson propagation might become very important~\cite{V04}.
We however postpone  study of these effects to a future work. At
sufficiently large baryon densities there might appear condensates
of some (quasi)Goldstone bosons, like $K^-$ and $\bar{K}^0$,
$\rho^{-}$, and might be $\eta$. To deal with stable ground state
we include a self-interaction of (quasi)Goldstone boson fields.
With this model we construct the EoS  as function of the
temperature and the baryon density and apply it in a broad
density-temperature region. Below for brevity we call thus
constructed model as the scaled hadron mass and couplings (SHMC)
model.

The paper is organized as follows. In sect. \ref{Lagr}  we
formulate the Lagrangian of the model. Then in sect. \ref{Endens}
the energy density for a system at finite baryon density and
temperature is constructed. In sect. \ref{Det} parameters of the
model are determined by fitting them to available data at zero
temperature or by exploiting the symmetry relations in case when
there are no data. Sects. \ref{EoS0} and \ref{EoST} are devoted to
evaluation of thermodynamic properties of the constructed EoS for
$T=0$ and $T\neq 0$, respectively. In sect. \ref{hic} we apply our
model to HIC. Some concluding remarks and perspectives are given
in sect. \ref{Concl}. Lengthy formulae for different terms of the
energy density and  a scheme how to calculate condensates, when
they occur in some $(T,n_B)$ range, are deferred to Appendices A
and B. Throughout the paper we use units $\hbar =c=1$.

\section{Lagrangian}\label{Lagr}
Within our model we present the Lagrangian density of the hadronic matter
as the sum of several terms:
 \be\label{math}  \mathcal{L}=\mathcal{L}_{\rm bar}+\mathcal{L}_{\rm mes}+\mathcal{L}_{\rm
   Gold}+\mathcal{L}_{\rm el}+\delta \mathcal{L}_{\rm vec}.\ee
Let us describe each term in (\ref{math}).

The  Lagrangian density for baryons interacting via $\sigma,\omega,\rho$
is as follows, cf. \cite{KV04},
 \be \mathcal{L}_{\rm bar} &=& \sum_{b\in \{b\}} \left[ \bar \Psi_b\,
\Big(i\,D\cdot
\gamma\Big)\, \Psi_b -m_b^*\, \bar\Psi_b\,\Psi_b +t_Q^b n_b V
\right] . \label{lagNn} \ee
The long derivative $D$
is given as follows
\be
 && D_\mu=\prt_\mu -i V \ t_Q^b \ \delta_{\mu 0}+i\,g_{\om b} \,{\chi}_\om
 \ \om_\mu+ ig_{\rho b}\, {\chi}_\rho\,
\vec{\rho}_\mu\,\vec{t}_b\,, \ee where $g_{\om b}$ and $g_{\rho
b}$ are  coupling constants and $\chi_\om$, $\chi_\rho$ are
coupling  scaling functions which will be determined below,
$\om^\mu=(\om_0,\vec {\om})$ and $\rho_\mu^a=(\rho_0^a,\vec
{\rho}^{\,a})$ are $\om$- and $\rho$-fields with $a=1,2,3$.
\begin{table}
\caption{Properties of the basic baryon set (masses are given in
MeV) } \label{tabl1}
\begin{tabular}{|l|rrrrrrrrrr|} \hline
  &
$p$&$n$&$\Delta^{++}$&$\Delta^{+}$&$\Delta^{0}$&$\Delta^{-}$&$\Lambda^0$&$\Sigma^+$
  &$\Sigma^0$&$\Sigma^-$ \\ \hline
$m_b$&938&938&1232&1232&1232&1232&1116&1193&1193&1193 \\
$t^3_b$&$\frac{1}{2}$&$-\frac{1}{2}$&$\frac{3}{2}$&$\frac{1}{2}$&$-\frac{1}{2}$
&$-\frac{3}{2}$&0&1&0&-1 \\
$t^Q_b$ &1&0&2&1&0&-1&0&1&0&-1 \\
$t^s_b$ &0&0&0&0&0&0&-1&-1&-1&-1  \\
$s_b$&$\frac{1}{2}$&$\frac{1}{2}$&$\frac{3}{2}$&$\frac{3}{2}$&$\frac{3}{2}$&$\frac{3}{2}$
&$\frac{1}{2}$&$\frac{1}{2}$&$\frac{1}{2}$&$\frac{1}{2}$ \\
\hline
\end{tabular}
\\[2mm] continuation \\[2mm]
\begin{tabular}{|l|rrrrrrrr|} \hline
  & $\Xi^0$&$\Xi^-$&$\Sigma^{*+}$&$\Sigma^{*0}$&$\Sigma^{*-}$
  & $\Xi^{*0}$&$\Xi^{*-}$&$\Omega^{-}$ \\ \hline
$m_b$&1318&1318&1385&1385&1385&1530&1530&1672 \\
$t^3_b$&$\frac{1}{2}$&$-\frac{1}{2}$&1&0&-1&$\frac{1}{2}$&$-\frac{1}{2}$&$-\frac{1}{2}$
\\
$t^Q_b$ &0&-1&1& 0&-1 &0&-1& -1\\
$t^s_b$ &-2&-2&-1&-1&-1&-2&-2&-3 \\
$s_b$&$\frac{1}{2}$&$\frac{1}{2}$&$\frac{1}{2}$&$\frac{1}{2}$&$\frac{1}{2}$&$\frac{1}{2}$
&$\frac{1}{2}$&$\frac{1}{2}$ \\ \hline
\end{tabular}
\\[3mm]
\end{table}
Here $\gamma_\mu$ are Dirac matrices, $n_b$ is the  particle (or
antiparticle) density of baryon species $b$, $V=-e\Phi_{\rm el}$
is the electric potential $\Phi_{\rm el}$ measured in the electron
charge units.

The baryon set $\{b\}$ that we use is presented in Table 1. In
reality masses of charged and neutral particles of the given
species are slightly different.  We ignore this difference that
allows us to use isospin invariance for $V=0$.

The $\sigma$-, $\omega$-, $\rho$-meson contributions to the
Lagrangian density \be \mathcal{L}_{\rm mes}=\sum_{m\in \{m\} }
\mathcal{L}_{m},\quad \mbox{with} \quad \{m\}=\sigma , \om, \rho
\ee render, respectively:
\be
 \mathcal{L}_\sigma &=&\frac{\prt^\mu
\sigma \ \prt_\mu \sigma}{2}-\frac{m_\sigma^{*2}\,
\sigma^2}{2}-{U}(\sigma);
  \nonumber\\
\mathcal{L}_\om &=&-\frac{\omega_{\mu\nu}\, \omega^{\mu\nu}}{4}
+\frac{m_\om^{*2}\, \om_\mu\om^\mu}{2}, \quad \omega_{\mu\nu}\,=
\partial_\mu \om_\nu -\partial_\nu \om_\mu ;
\label{lagORn}
\ee
\be
\mathcal{L}_{\rho} &=& \mathcal{L}_{\rho^0}+\mathcal{L}_{\rho^{\pm}} ;\nonumber\\
\mathcal{L}_{\rho^0} &=&\frac{\prt^\mu \rho_0^3 \prt_\mu
\rho_0^3}{2}+\frac{1}{2}m_\rho^{*2}  (\rho_0^3 )^2 ,\\
\mathcal{L}_{\rho^{\pm}} &=&[(i\prt^0
  -V+g_\rho \ \chi_\rho^{'} \ \rho_0^{(3)} ) \ \rho_{ch}^- ] \ [(i\prt^0
  -V+g_\rho \ \chi_\rho^{'} \ \rho_0^{(3)} ) \ \rho_{ch}^-]^{\dagger}\nonumber\\
&-&|\nabla\rho_{ch}^- |^2-m_\rho^{*2}  |\rho_{ch}^- |^2 ,\quad
\rho_{ch}^- =(\rho_1 -i\rho_2 )/\sqrt{2} .
\ee
Here
$g_\rho$ coupling is responsible for the self-interaction of the
charged and neutral species,  $\chi_\rho^{'}$ is the scaling function.
 Non-Abelian $\rho$-$\rho$
interaction with $g_\rho =g_{\rho N}$ is motivated by the hidden
local symmetry approach, cf. \cite{bando83}, where $\rho$-meson is
introduced as a non-Abelian gauge boson. Nevertheless this
possibility is  often disregarded and one uses the simplest form
with $g_\rho =0$, cf. \cite{HFN}. In a sufficiently dense
asymmetric nuclear matter the presence of the self-interaction may
result in appearance of the charged $\rho$-meson condensate
characterized by non-zero $\rho^{-}$ mean field instead of
$\rho^0$ one, cf. \cite{KV04,V97}.

Following \cite{KV04} we use the $\sigma$-field dependent
effective masses of baryons \be \label{bar-m}
{m_b^*}/{m_b}=\Phi_b(\chi_\sigma  \sigma)= 1 -g_{\sigma b} \
\chi_{\sigma} \ \sigma /m_b \,, \; b\in\{b\} \ee with the baryon
set $\{b\}$ defined in Table 1 and mass terms of the mean fields
are
\be
\label{bar-m1} {m_m^*}/{m_m}&=&|\Phi_m (\chi_\sigma
\sigma)|\,, \quad \{m\}=\sigma,\om,\rho\,,
\ee
where $g_{\sigma
b}$ are $\sigma b$-coupling constants. We have introduced the
absolute value of $\Phi_m (\chi_\sigma \sigma)$ to indicate that
only $(m_m^* )^2$ mass terms, as they appear in the Lagrangian,
have physical meaning. This observation is important to interpret
the situation when $\Phi_m (\chi_\sigma \ \sigma)$ becomes
negative, see sect. \ref{EoST}  below.

For the sake of simplicity we scale all  couplings $g_{\sigma b}$
by a single scaling function $\chi_{\sigma}(\sigma )$,
and all $g_{\om b}$, $g_{\rho b}$ by $\chi_{\om}(\sigma )$ and
$\chi_{\rho}(\sigma )$ scaling functions, respectively. Thus all
scaling functions depend only on $\sigma $. The idea behind that
is as follows. The $\sigma $ field can be expressed in terms of
the $ud$ quark condensate. The change of effective hadron masses
and couplings is associated namely with  modification of the quark
condensate in matter.  Thus we consider the $\sigma$ field as an
order parameter. The $\sigma$ excitations are then treated as
fluctuations around the mean value of the order parameter.
Similarly long-scale fluctuations  are treated in the Landau
phenomenological theory of phase transitions.

The dimensionless scaling functions $\Phi_b$ and $\Phi_m$, as well
as the coupling scaling functions $\chi_m$ depend on the scalar
field in the combination $\chi_\sigma(\sigma) \ \sigma$. Therefore
for further convenience we introduce the variable \be\label{f}
f=g_{\sigma N} \ \chi_\sigma \ \sigma/m_N\,. \ee Following
\cite{KV04} we assume an approximate validity of the Brown-Rho
scaling ansatz in the simplest form \be \label{Br-sc}\Phi =\Phi_N
=\Phi_\sigma =\Phi_\om =\Phi_\rho =
 1-f .
 \ee

 We keep the
standard form for the non-linear self-interaction term  (potential
$U$) of RMF models, but now in terms of the new variable $f$, and
using (\ref{f}) it can be rewritten as follows:
\be
U&=&m_N^4
(\frac{b}{3}\,f^3 +\frac{c}{4}\,f^4 ) =\frac{b m_N\,(g_{\sigma
N}\,\chi_\sigma \, \sigma)^3} {3} +\,\frac{c (g_{\sigma
N}\,\chi_\sigma \, \sigma)^4} {4}\,. \label{Unew}
\ee
Two
additional parameters, $b$ and $c$, allow us to accommodate
realistic values of the nuclear compressibility and the effective
nucleon mass at the saturation density. An extra attention should
be paid to the fact that the coefficient $c$ must be positive to
deal with the stable ground state.

The contribution of the electric field $\mathcal{L}_{\rm el}$ to
the Lagrangian density is:
\be
\mathcal{L}_{\rm el} =\frac{1}{8\pi
e^2}(\nabla V)^2~.
\ee
Coulomb effects are  responsible for a
deviation of the low-momentum $\pi^+ /\pi^-$ rates  from unity in
HIC of isospin-symmetric nuclei, and for some other effects. It is
important to include the Coulomb term for the description of mixed
phases in dense neutron star matter, cf. \cite{VYT,MVTC05}.

There are mean-field solutions of the Lagrangian $\mathcal{L}_{\rm
bar} +\mathcal{L}_{\rm mes} +\mathcal{L}_{\rm el} \equiv
\sum_{b\in \{b\}}\mathcal{L}_b +\sum_{m\in \{m\}}\mathcal{L}_m
+\mathcal{L}_{\rm el}$. To these terms we add the Lagrangian
density
\be
\label{Gold} \mathcal{L}_{\rm Gold}=\sum_{g\in
\{g\}}\mathcal{L}_{g},\,\,\,\quad \{g\} =\pi^{\pm ,0} (138);K^{\pm
,0} , \bar{K}^0 (495); \eta (547).
\ee

These particles  are often treated as (quasi)Goldstone ("$\rm
Gold$") bosons within the chiral SU(3) symmetrical models.
Therefore we may not to scale their masses and couplings, as we
have done for $\{m\}=\sigma ,\om ,\rho $. At rather small baryon
densities there are no mean-field solutions of equations of motion
which follow from $\mathcal{L}_{\rm Gold}$. Such solutions may
however arise at sufficiently large baryon densities signalizing
on condensations of these fields. On the other hand, we observe,
cf. \cite{KV04}, that for the case of spatially homogeneous system
equations for mean fields and thus mean-field solutions do not
change if we replace $\sigma$-, $\om$-, $\rho$-fields by the
scaled fields $\chi_{\sigma }\sigma$, $\chi_{\om }\om$ and
$\chi_{\rho }\rho$ provided $\Phi_b =\Phi_m =\chi_{m}$, and
$\chi_{\rho}^{\prime}=\chi_{\rho}^2$. If we wish to extend this
symmetry to the case when Goldstones are  included, in addition to
scaling of masses we should scale couplings, $g^*_{mg} =g_{mg} \
\chi_{m}$. Below we will test both possibilities $g^*_{mg}
=g_{mg}$ and $g^*_{mg} =g_{mg} \ \chi_{m}$, and refer to them as
versions without and with scaling, respectively.

 The contribution of the pion Lagrangian density $\mathcal{L}_{\pi}$
 into Eq.~(\ref{Gold}) is given by
\be
\mathcal{L}_{\pi}&=&\mathcal{L}_{\pi^0}+\mathcal{L}_{\pi^{\pm}},\\
\mathcal{L}_{\pi^0} &=&\frac{\prt^\mu \pi^0 \prt_\mu
\pi^0}{2}-\frac{m_{\pi}^{*2}\, (\pi^0 )^2}{2},\quad
m_{\pi}^{*}=m_{\pi}-g_{\sigma \pi}^*\sigma ;  \\
\mathcal{L}_{\pi^{\pm}}&=&(i\prt_0 -V+g_{\omega \pi}^* \omega_0
+g_{\rho \pi}^*\rho_0^3 )\pi^{-} \ [(i\prt_0
-V+g_{\omega\pi}^*\omega_0
+g_{\rho \pi}^*\rho_0^3 )\pi^{-}]^{\dagger} \nonumber \\
&-&|\nabla \pi^{-}|^2 -m_{\pi}^{*2}|\pi^{-}|^2~.
 \ee
 The kaon Lagrangian density contributes as
\be
\mathcal{L}_{K}&=& [(\prt_0 -i\hat{q}_K V+ig_{\omega
K}^*\omega_0
  +ig_{\rho K}^*\tau_3 \rho_0^3 )^{\dagger}K^{\dagger}]\nonumber\\
&\times&[(\prt_0 -i\hat{q}_K V +ig_{\omega K}^*\omega_0
  +ig_{\rho K}^*\tau_3\rho_0^3 )K]\\
  &-&\nabla K^{\dagger}\nabla K -m_{K}^{*2}K^{\dagger} K , \quad
  m_{K}^{*}=m_{K}-g_{\sigma K}^*\sigma , \quad \hat{q}_K =\frac{1}{2}
  (1+\tau_3)~, \nonumber
\ee
with $K=(K,{K}^0)$, $\tau_3$ is the Pauli matrix.

We follow \cite{ZPLN} and present the
 Lagrangian density of the $\eta$ meson as
\be \label{Leta} \mathcal{L}_\eta &=&
\frac{1}{2}\partial^{\mu}\eta \ \partial_{\mu}\eta -
\frac{1}{2}\left(m_{\eta}^2 -\sum_{b\in\{b\}} \frac{\Sigma_{\eta
b}}{f^2_{\pi}} \bar\Psi_b
\,\Psi_b \right)\eta^2 \nonumber\\
&+&
\frac{1}{2}\sum_{b\in\{b\}}\frac{\kappa_{\eta b}}{f^2_{\pi}}\bar\Psi_b
\,\Psi_b \
\partial^{\mu}\eta \ \partial_{\mu}\eta .
 \ee
Here $f_{\pi}=93$~MeV is the pion decay constant. For nucleons the
constants $\Sigma_{\eta N}$ and $\kappa_{\eta N}$ are estimated
from the scattering data: $\Sigma_{\eta N}\simeq 280\pm 130~$MeV
and $\kappa_{\eta N} \simeq 0.40\pm 0.08$~fm. Other parameters in
(\ref{Leta}) are not known. Ref. \cite{ZPLN} used a large value
for the $KN$-sigma term, $\Sigma_{K N}$. Below we argue for a
smaller value of  $\Sigma_{K N}$. Therefore we will test a
sensitivity of the $\eta$-description to the coupling variations.

When the total Lagrangian is known, one can derive equations of
motion for every field. Even for low
baryon density, equations of motion for $\sigma$, $\om$
and $\rho$ and $V$ allow  mean-field solutions  $\sigma_0$, $\om_0$,
$\rho^3_0$, and $V_0$. Therefore we
use:
\be
\sigma\equiv \sigma_0; \quad \om_{\mu}=\om_0 \
\delta_{\om 0}; \quad \rho_\mu^a = R_0 \ \delta_{a
  3} \ \delta_{\mu 0}; \quad V\equiv V_0 .
  \ee
 Only for isotopically asymmetric matter ($N\neq Z$) we have $R_0\neq 0$.
As we have mentioned, if the baryon density increases above a
critical value and $N\neq Z$, there may appear another solution
with $R_0 =0$ but with a non-zero solution for the charged
$\rho$-meson mean field,
 $\rho_{ch}^- \neq 0$, cf.~\cite{KV04,V97}. For the sake of simplicity
we disregard such a possibility in the present work.

Similarly to the case of self-interacting $\rho$ meson fields,
there exist higher order terms  in $\mathcal{L}_{\rm Gold}$
describing self-interaction of the fields. Using approximate SU(3)
theory these terms can be presented  as \be\label{self}
\mathcal{L}^{\rm int}_{\rm Gold} =\lambda \sum_g (\vec{\phi}^{\,2}
)^2 /4 ;\quad \vec{\phi}= (\pi_1 , \pi_2 , \pi_3 ; K_1 , K_2 , K_3
,K_4 ; \eta ) \ee with a positive self-interaction coupling
constant $\lambda =const \sim 1$ and redefined  fields
$\pi^{\pm}=(\pi_1 \pm i\pi_2 )/\sqrt{2}$, $\pi^0 =\pi_3$,
$K^{\pm}=(K_1 \pm i K_2 )/\sqrt{2}$, $K^{0}=(K_1 +i K_2
)/\sqrt{2}$, $\bar{K}^{0}=(K_1 -i K_2 )/\sqrt{2}$. Eq.
(\ref{self}) has the simplest form although we could use
self-interaction terms with different couplings for different
particle species.

The remaining terms $\delta\mathcal{L}_{\rm vec}$ in the
Lagrangian density (\ref{math}) are due to the vector mesons $K^*$
and $\varphi$, and the glueball-like state $\eta^{'}$. One could
treat $\varphi$ like $\om$ with similar scaling of the effective
masses and couplings. Due to a high value of strange quark mass
one can expect that $K^*$ and $\varphi$ couplings are less than
those for $\om$. Since little is known about interactions of these
particles and not to complicate further consideration by
introducing unknown parameters we consider $K^*$ and  $\varphi$ as
free particles in present work. To our knowledge there is no
information about values of $\eta^{'}b$-couplings. Therefore,
being conservative, we put them zero treating $\eta^{'}$ also as a
free particle.

As we have mentioned,  condensates of some (quasi)Goldstone fields
may appear at some specific conditions. In this case their
equations of motion acquire mean-field solutions, as those for
$\sigma$, $\om$ and $\rho$. In such cases for neutral fields (the
strangeness and electric charge being zero) the stability of the
ground state is provided only due to presence of the
self-interaction, see Eq.~(\ref{self}).

To single out quasiparticles (excitations) from mean fields, in
the Lagrangian $\mathcal{L}_{\rm mes}$ one should do replacements
$\om_0 =\om_0^{\rm cl}(\sigma ) +\om^{\prime} $, $R_0 =R_0^{\rm
cl}(\sigma ) +R_0^{\; \prime}$, $\vec{\phi} =\vec{\phi}^{\rm
cl}(\sigma ) +\vec{\phi}^{\; \prime}$,
 $\vec{\om}=\vec{\om}^{\; '}$ and $\vec{\rho}=\vec{\rho}^{\; '}$.
Here $\om_0^{\rm cl}$, $R_0^{\rm cl}$, $\vec{\phi}^{\rm cl}$ are
the mean (classical) field variables and
$\om_\mu^{\prime}$, $(\rho^{\prime}_0 )^{\mu}$,
$(\rho^{\prime}_{\pm} ) ^{\mu}$, $\vec{\phi}^{\prime}$ are
responsible for  new excitations. Then we expand the Lagrangian
density retaining only quadratic terms in the fields of
excitations. The coefficients at non-derivative quadratic terms
are read as squared masses of excitations. We recognize that
effective masses of the zero and spatial components of vector
fields are equal and the gauge conditions $\partial^\mu \om_\mu
=0$, $\partial^\mu \rho_\mu^a =0$ are fulfilled.

Equations of motion for mean fields  and for excitations are
obtained by the variation of the total action. If mean-field terms
are rather large and excitation contributions are small,  one may
disregard the excitation terms in equations of motion for the mean
fields and neglect self-interactions of excitations. However one
should keep  interactions of excitations with mean fields in
equations of motion for excitations and in their thermodynamic
quantities.

Since there are no experimental indications of
 condensation of (quasi)Goldstone bosons in the regimes
of HIC, we will focus our  further discussion on the case
 when  condensates do not occur, paying a special attention to
situations when condensation is possible.

We checked that minimization of the energy density with respect to
the $\sigma$ field produces an equation of motion for this field
being in agreement with the Lagrange equation after its Gibbs
averaging. Thus to find the squared effective mass of the $\sigma$
excitation (i.e. of the fluctuation of the order parameter), we
take $\om_0 (\sigma )$, $R_0 (\sigma )$, plug them in the energy
density, put $\sigma =\sigma^{\rm cl}+\sigma^{'}$, and find the
second derivative of the energy density in respect with
$\sigma^{\rm cl}$, see Eq. (\ref{sigm-mass}) in Appendix A.

\section{Energy density at finite temperature}\label{Endens}
Let us assume that the system volume is sufficiently  large and
surface effects may be disregarded. Thus  only spatially
homogeneous RMF solutions of the equations of motion are
considered. To simplify expressions we will  treat all quantities
in the rest frame. Generalization to the arbitrary moving inertial
frame is obvious.

 The thermodynamic potential density $\Omega$, pressure $P$, free
 energy density $F$, energy density $E$ and entropy density are
 related as
\be\label{presmu}
&&E=F+TS, \quad
  F[f,\om_0 ,R_0 ,T] =\sum_{i} \mu_i n_i +\Omega\,,\quad \Omega =-P,\\
  &&\mu_i=\frac{\prt F}{\prt n_i}
\,.
  \ee
Summation index $i$ runs over all particle species; $n_i$ are
particle densities, see Eq. (\ref{dens}). Chemical potentials
$\mu_i$ enter  Green functions in the standard gauge combinations
$\varepsilon_i +\mu_i$.

The  energy density can be presented as the sum of the mean
$\sigma$-, $\omega$-, $\rho$-field contributions as well as
contributions of baryons and all meson excitations. So we have
\be
\label{Efun} E[f,\om_0 , R_0 ,T] &=& \sum_{b\in\{b\}} E_b
[f,T]+\sum_{m\in\{m\}}
E^{\rm MF}_m [f,
\om_0 ,R_0 ,T]\nonumber\\
&+&E_{\rm bos. ex.}[f,\om_0 ,R_0 ,T]\equiv E_{\rm MF}+ E_{\rm bos.
ex.}~.
\ee
 The first two sums (resulting in the term $E_{\rm MF}$) are included
 in every RMF model but with smaller set $\{b\}$, whereas the boson
 excitation term $E_{\rm bos. ex.}$ is obtained here  beyond
the scope of the RMF approximation.

Although all terms are functions only of $f$ and $T$, we will
present them also as functions of $\om_0$ and $R_0$ in such a way
that the values of the  $\om_0 (f)$ and $R_0 (f)$ mean fields can
be found by minimization of the energy at fixed $f$. Then $\om_0
(f)$ and $R_0 (f)$ are plugged in the energy density functional
that becomes function of $f$ only. So the equilibrium value of $f$
can be found by subsequent minimization of the energy in this
field.

Since $E_{\rm bos.
ex.}[f,\om_0,R_0,T]$ depends on the mean fields, its minimization
 produces extra terms in the mean-field equations. Within the
approximation of rarefied gas of excitations used in this work, we
treat excitations perturbatively thus omitting these extra terms.
Therefore, we assume that   $E_{\rm bos. part} =E_{\rm bos.
part}[f^{\rm MF},\om_0^{\rm MF} ,R_0^{\rm MF} ,T]$, where $f^{\rm
MF},\om_0^{\rm MF} ,R_0^{\rm MF}$ are found by minimization of the
energy $E=E_{\rm MF}$, i.e. without inclusion of the boson
excitation term. Thus our equations of motion for mean fields are:
\be
\frac{\prt}{\prt \om_0 }\,E_{\rm MF} [f,\om_0 ,T ]=0\,~, \quad
\frac{\prt}{\prt R_0 }\,E_{\rm MF} [f,R_0 ,T]=0\,\label{extreme}
\ee
 and
 \be \frac{d}{d f}\,E_{\rm MF}[f,\om_0 (f), R_0 (f),
T]=0\,. \label{extremef} \ee At variation of the energy density
one should not vary over the particle occupation numbers. Note
that if the interaction of excitations were included within the
self-consistent Hartree approximation, in Eqs. (\ref{extreme}),
(\ref{extremef}), we would minimize the total energy $E$ rather
than $E_{\rm MF}$.  It would however additionally complicate the
solution of the problem. We postpone the study of this possible
model generalization for a future work.

It is noteworthy that to obtain equations of motion, instead of
the energy density one could vary the thermodynamic potential
$\Omega$, cf. \cite{MVTC05}.

 At the resonance peak   the vacuum $\Delta$-isobar mass
width   is $\Gamma_{\Delta}^{\rm max}\simeq 115$~MeV. In reality
$\Gamma_{\Delta}$ is the temperature-, density- and
energy-momentum-dependent quantity. For low $\Delta$-energies the
width is much less than $\Gamma_{\Delta}^{\rm max}$. A typical
$\Delta$ energy is $\om_\Delta-m_\Delta^*\sim T$. Thus for low
temperature, $T\lsim \epsilon_F$ ($\epsilon_F$ is the nucleon
Fermi energy) the effective value of the $\Delta$-width is
significantly less than $\Gamma_{\Delta}^{\rm max}$. At these
temperatures the quasiparticle approximation does not work for
$\Delta$'s but their contribution to thermodynamic quantities is
small. When the temperature is  $\gsim m_\pi$  there appears
essential temperature contribution to the width and the resonance
becomes broader \cite{VS91}. $\Delta$'s essentially contribute to
thermodynamic quantities. Only for temperatures $T\gsim
\Gamma_{\Delta}^{\rm max}(T)$ the quasiparticle approximation
becomes a reasonable approximation.

In reality $\rho$- and $\sigma$-mesons also have rather broad
widths.  The observed enhancement of the dilepton production at
CERN, {in particular in the recent NA60
experiment~\cite{NA60} on $\mu^+\mu^-$ production,} can be
explained by significant broadening of the $\rho$ in matter
\cite{RW}, though decreasing of the $\rho$ mass could also help in
explanation of the data \cite{TS}\footnote{{As
demonstrated in \cite{TS} the calculated large mass shift is
mainly caused by the assumed temperature dependence of the
in-medium mass.
Inclusion of this temperature dependence
modifies the scaling hypothesis originally claimed by Brown and
Rho. Some arguments on what the proper
mass-scaling  predicts for dilepton production in HIC, e.g. NA60,
were given in \cite{BR06}.}}. Besides, the $\rho$ width might
increase with further decrease of its effective mass \cite{RZM}.
Also particles which have no widths in vacuum like nucleons
acquire the widths in matter due to collisional broadening. Their
widths grow with   the temperature increase, cf. \cite{V04}. As we
have argued above in case with $\Delta$ isobars, the quasiparticle
approximation may become a reasonable approximation at
sufficiently high temperature, if $T\gsim \Gamma (T)$.

 The problem becomes much more involved, if one tries to
treat particle width effects consistently. Therefore, to simplify
consideration we use the quasiparticle approximation in the
present work for all particle species in the whole temperature and
baryon density range of our interest.

Now let us subsequently  consider all energy terms in Eq.
(\ref{Efun}).

\subsection{The baryon contribution}
The  contribution of the given baryon species $b$ to the energy
density is as follows
\be\label{Eb}
 E_b [f,T]&=& (2 s_b +1)\intop_0^{\infty}\frac{\rmd p \
p^2}{2\pi^2}\,(f_b +\bar{f}_b )\sqrt{m_b^{*2}(f)+p^2}-t^Q_b \ n_b \
 (V+\mu_{\rm ch})  \nonumber\\
&+&t^Q_b \ \bar{n}_b \ (V+\mu_{\rm ch}), \quad \quad p=|\vec{p}|.
 \ee
The spin factor $s_b =1/2$ for $N$ and hyperons, while
$s_b =3/2$ for the $\Delta$-resonance, see Table 1.
The Fermi-particle (baryon) occupations,

\be\label{oc} f_b  &=&\frac{1}{\exp[(\sqrt{m_b^{*2}+p^2}-\mu_b^*)/T]+1},\\
\bar{f}_b
&=&\frac{1}{\mbox{exp}[(\sqrt{m_b^{*2}+p^2}+\mu_b^*)/T]+1}~,
\ee
depend on the gauge-shifted values of the chemical potentials
 \be \label{Tmu}
 \mu_b^* =t_b  \ \mu_{{\rm bar}}+t^s_b \ \mu_{{\rm
str}}+t^Q_b (\mu_{\rm ch} +V) -g_{\om b} \ \chi_{\om}\om_0 - t^3_b
\ g_{\rho b} \ \chi_{\rho} \ R_0 ~.
\ee

The baryon chemical potential of the $b$ species is $\mu_b =t_b \
\mu_{{\rm bar}}$, and the corresponding strangeness term is
$\mu^{s}_b =t^s_b \ \mu_{{\rm str}}$. As is seen, the electrical
potential $V\rightarrow V+\mu_{\rm ch}$ is shifted by the charge
chemical potential $\mu_{\rm ch}$ related to the isospin
composition of the system. Suppressing Coulomb effects one drops
out  the shifted value of $V$. Sometimes instead of $\mu_{\rm ch}$
one introduces the isospin chemical potential, cf.~\cite{TDGGL}.

\subsection{Mean-field contribution}
It is convenient to  introduce the coupling ratios
 \be\label{rat}
 x_{mb}=g_{mb} /g_{mN}, \,\,\, \{ m\}=\sigma ,\om ,\rho ,
 \ee
and, instead of $\chi_m$, another variables
\be
\eta_{m}(f)={\Phi_m^2(f)}/{{\chi}_m^2(f)}\,,
\label{eta}
\ee
since
the energy density depends namely on such combinations rather than
on $\Phi_m$ and $\chi_m$ separately.

Using these new variables the contribution of  mean fields to the
energy density is given as follows:

 \be \label{Esig}E^{\rm MF}_\sigma [f]=
\frac{m_N^4\,f^2}{2\, C_\sigma^2}\, \eta_{\sigma}(f) +{U}(f) ,\ee

\be
E^{\rm MF}_\om [f,\om_0 ] &=& \frac{C_\om^2\, (\sum_{b\in \{b\}}x_{\om
b} \ (n_b -\bar{n}_b ) )^2}{2\, m_N^2\, \eta_{\rm \omega}(f)}
\nonumber\\
&-&\frac{m_N^2\eta_\om(f)}{2\,C_\om^2}\left[g_{\om
N}\,{\chi}_\om\,\om_0 - \frac{C_\om^2 \ (\sum_{b\in \{b\}}x_{\om
b} \ (n_b -\bar{n}_b))}{\,m_N^2\,\eta_\om (f)}\right]^2\,.
\label{ome}
\ee
The net baryon density  is defined as follows

\be &&\sum_{b\in \{b\}}(n_b -\bar{n}_b)\equiv n_B ~, \ee
where the
partial baryon and antibaryon densities for the species $b$ are
 \be
n_b =(2s_b +1)\intop_0^{\infty}\frac{\rmd p \ p^2}{2\pi^2}\,f_b~,
\quad \bar{n}_b =(2s_b +1)\intop_0^{\infty}\frac{\rmd p \
p^2}{2\pi^2}\,\bar{f}_b~.
\ee

 Renormalized constants are
\be
C_m=\frac{m_N \ g_{mN}}{m_m}
. \label{Cm}
\ee
Values of the parameters used will be specified
in sect. 4 below. Similarly, for the $\rho$ mean-field
contribution we have

\be \label{Erho}E^{\rm MF}_{\rho}[f,R_0 ,T ]= \frac{C_\rho^2 \
(n^t_{B})^{2}}{8\,m^2_N\, \eta_\rho(f)}
-\frac{m_N^2 \ \eta_\rho(f)}{2\,C_\rho^2}\left[g_{\rho
N}\,{\chi}_\rho \ R_0 - \frac{C_\rho^2 \ n^t_{B}}
{2\,m_N^2\,\eta_\rho (f)}\right]^2\!  . \ee

The isotopic charge density in the baryon sector is given by
 \be
 n^t_{B} =2 \sum_{b\in\{b\}} \ t_b^3 (n_b - \bar{n}_b ) \ x_{\rho b}~.
 \ee
As one can see,  the isovector baryon density $n^t_{B}$ plays the
role of the source for the $\rho$-meson field
$\rho_0^{(3)}=R_0$\,. Therefore for the iso-symmetrical matter
($N=Z$) one has $n_t^B =0$ and $E^{\rm MF}_\rho =0$\,.

The net density of strange baryons and mesons reads
 \be\label{nt}
 n_{\rm
 str}=\sum_{b\in\{b\}}t_b^s \ (n_b -\bar{n}_b )&-&
n_{K^-}-n_{\bar{K}^0}-n_{K^{*-}}-n_{\bar{K}^{*0}}\nonumber\\
&+& n_{K^+}+n_{K^0}+n_{K^{*+}}+n_{K^{*0}}~.
\ee
We assume that all
strange particles are trapped inside the fireball till the
freeze-out. Therefore the total strangeness is zero. In this paper
we do not consider the possibility of a mixed phase: Strange
clusters surrounded by normal matter. This possibility arises
since the charge (electric charge, baryon charge, strangeness,
etc) can be conserved only globally rather than
locally~\cite{G92}. We put locally $n_{\rm
 str}=0$. Then this condition determines the value of the strangeness chemical
potential $\mu_{\rm str}$.

Similarly, we may introduce the electric charge density \be n_{\rm
ch} = \sum_{b\in\{b\}} t_b^{Q}(n_b -\bar{n}_b )&+&
n_{\pi^+}+n_{K^+}+n_{\rho^+}+n_{K^{*+}}\nonumber\\
&-&n_{\pi^-}-n_{K^-}-n_{\rho^-}-n_{K^{*-}}\, ,
\ee
assuming that
it is conserved locally.  The quantity  $n_{\rm
  ch} =(Z/A)n_{B}$
determines the value of the charged chemical potential $\mu_{\rm ch}$.

Our SHMC RMF energy density functional   depends on \emph{four}
particular combinations of the functions,
$\eta_{\sigma,\rho,\om}(f)$ and $U(f)$. Note that the dependence
on the scaling function $\eta_{\sigma}$ can always  be presented
as a part of the new potential $U$ obtained by means of the
replacement $U\to U+\frac{m_N^4\,f^2}{2\, C_\sigma^2}\,
(1-\eta_{\sigma}(f))$\,, and vise versa, so the potential $U$ can
be absorbed in the new quantity $\eta_{\sigma}$. Thus actually
only \emph{three} independent functions enter the energy density
functional. Eq.~(\ref{Efun}) together with
Eqs.~(\ref{Eb}),~(\ref{Esig}),~(\ref{ome}),~(\ref{Erho})
demonstrate explicitly equivalence of mean-field Lagrangians  for
constant fields with various parameters if they correspond to the
same functions $\eta_{\rho,\om}(f)$ and $\eta_{\sigma}$ (either
$U(f)$), with the field $f$ related to the scalar field $\sigma$
through Eq.~(\ref{f}).

\subsection{Bosonic excitations}
To find the total energy (\ref{Efun})
 one should yet define  the contribution of bosonic
excitations.
The  energy density of boson
excitations is the sum of partial contributions
\be
\label{exden} E_{\rm bos.ex}[f,\om_0 ,R_0
,T]&=&E_{\sigma}^{\rm
part}+E_{\om}^{\rm part}+E_{\rho}^{\rm part}+E_\pi^{\rm part}\nonumber\\
&+&E_K^{\rm part}+ E_{\eta}^{\rm part} +E_{K^*}^{\rm part}
+E_{\eta^{\prime}}^{\rm part} +E_{\phi}^{\rm part}~.
\ee
Explicit
expressions for the partial contributions can be found in
Appendix~A.

As was mentioned, in the present paper we  consider {\em{a
non-interacting gas of excitations}}.  Thus, in order to get
$E_{\sigma}^{\rm part}$ we expand $E_{\rm MF}[\sigma ,\omega_0
(\sigma ) ,R_0 (\sigma ), T]$ in Eq. (\ref{Efun})  in $\delta
\sigma =\sigma -\sigma^{\rm cl}$. Linear term in $\delta \sigma$
does not contribute due to subsequent requirement of the energy
minimum in $\sigma^{\rm cl}$. From the second order term we
extract
\be
\label{spa} (m_{\sigma}^{\rm part*})^2 \equiv {\prt^2
E_{\rm MF}[\sigma ,\omega_0 (\sigma ),R_0 (\sigma ), T]}/{\prt
\sigma^2}.
\ee
Here particle occupation are not varied as in
(\ref{extremef}). In the gas approximation we may drop the higher
order terms in $\delta \sigma$. Effective masses of $\om$ and
$\rho$ prove to be the same as those follow from the mean-field
mass terms
\be
\label{omr} m_\om^{\rm part
  *} =m_\om |\Phi_\om (f)|, \quad m_\rho^{\rm part
  *} =m_\rho |\Phi_\om (f)|~.
\ee As have been mentioned, the simplifying ansatz (\ref{Br-sc})
is used in present work.

At certain conditions Bose condensates of some boson species may
occur. Below we will demonstrate that our choices of couplings  do
not allow for Bose condensates of excitations in the
temperature-density region which we will discuss in application to
HIC. However condensates may appear for other possible choices of
couplings and if a broader density-temperature interval is
considered. Explanation how to include Bose condensates if they
appear is given in Appendix~B.

\section{Choice of SHMC model parameters and scaling functions}\label{Det}

Parameters of the RMF model, $C_\sigma$, $C_\om$, $C_\rho$, and
the self-interaction potential $U$, are to be adjusted to
reproduce the nuclear matter properties at the saturation for
$T=0$. Usually they are fixed by values of the binding energy
$e_{\rm bind}$, nuclear saturation density $n_B=n_0$ and symmetry energy
coefficient $a_{\rm sym}$, which are known within some error bars.
We will use the same basic input parameters as in
Ref.~\cite{KV04}:
\be &&n_0 =0.16~\mbox{fm}^{-3}, \quad e_{\rm
bind}=-16~\mbox{MeV}, \quad a_{\rm sym}(n_0)=32~\mbox{MeV}.
\label{param}
\ee

The saturation baryon density and the binding energy are related
as
\be \frac{\prt E[n_B;f]}{\prt n_B}\Bigg|_{n_0,f(n_0)}=
\frac{1}{n_0}\,E[n_0;f(n_0)]=m_N+e_{\rm bind}\,, \label{satur}
\ee
and the compressibility modulus is given by
\be K=9
\,n_0\left[\frac{\prt^2 E}{\prt n_B^2}\Bigg|_{n_0,f(n_0)}- \left(
\frac{\prt^2 E}{\prt n_B\, \prt f}\Bigg|_{n_0,f(n_0)}\right)^2\,
\left[\frac{\prt^2 E}{\prt
f^2}\Bigg|_{n_0,f(n_0)}\right]^{-1}\right]\,. \label{kmod}
\ee

Here $f(n_0)$ is a solution of Eq.~(\ref{extremef}) at the density
$n_p=n_n=n_0/2$\,. The parameter $C_\rho$ is determined from the
symmetry energy coefficient of the nuclear matter
\be
a_{\rm
sym}(n_B)&=&\frac{n_B}{8}\frac{\prt^2}{\prt n_p^2} E \left(
n_B-n_p,n_p\right)\Big|_{n_p=n/2} \nonumber \\ &=&
\frac{C_\rho^2\, n_B}{8 m_N^2\, \eta_\rho}+\frac{\pi^2\, n_B}{4\,
p_F\,
 \sqrt{m_N^{*2}+p_{\rm FN}^2}}\,,
\label{symen}
\ee
$p_{\rm FN}$ being the nucleon Fermi momentum in
iso-symmetrical matter ($N=Z$).

We use the modified Walecka model with a non-universal scaling of
masses and couplings (referred as the MW(n.u) model in~\cite{KV04}
and as the KVOR model in~\cite{Army}). This model matches the
Urbana-Argonne EoS (A18+$\delta v$+UIX*)~\cite{APR98} for the
baryon densities below $4n_0$ at $T=0$, which correctly reproduces
the maximal neutron star mass $M_{\rm max}\simeq 2~M_{\odot}$ and
gives sufficiently large threshold density for the direct Urca
reaction $n_{\rm crit}^{\rm DU}$ to be in agreement with the
neutron star cooling phenomenology \cite{BGV}. The Urbana-Argone
(A18+$\delta v$+UIX*) EoS is derived within a microscopical
variational theory of nuclear matter. It employs the
non-relativistic paired $NN$ potential extracted from the analysis
of the scattering data, includes the boost $v^2/c^2$-order
corrections and incorporates a three nucleon interaction. However
due to using of the non-relativistic potential the Urbana-Argone
EoS violates causality for $n_B\gsim 4n_0$ and $T=0$. In
Ref.~\cite{HHJ}  the A18+$\delta v$+UIX* EoS is fitted for $n\lsim
4n_0$ but the causality problem is solved for higher densities. We
call this modification as the HHJ model. Parameters of our SHMC
(KVOR-based) EoS are fitted in such a way that the energy
(including the symmetry energy) and the pressure are very close to
those for A18+$\delta v$+UIX* (and HHJ) for $n_B< 4n_0$ and $T=0$
both for the $Z=0$ and $N=Z$ cases. Since our EoS is based on RMF
calculations, no causality problem arises.

An appropriate behavior of the EoS for $T=0$ is obtained with the
scaling factors introduced in the same way as in~\cite{KV04}

\be\label{etnun}
\eta_\sigma=\frac{\Phi^2_\sigma}{\chi^2_\sigma}=1\,, \ \eta_\om
(f)=\frac{1+z\,f(n_0)}{1+z\,f}\,, \ \ee \be\label{etnun1}\eta_\rho
(f)=\frac{\eta_\om (f)}{\eta_\om
(f)+4\,\frac{C_\om^2}{C_\rho^2}\,(\eta_\om (f)-1)}, \ee with $z$
as a parameter. If one puts $z=0$, the standard RMF version
without $\sigma$ scaling is covered. Effective nucleon mass and
the compressibility coefficient are
\be
m_N^{*}(n_0)/m_N =0.805,
\quad K=275~\mbox{MeV}. \label{MWnu-inp}
\ee
Note that, if we
chose a smaller values of $m_N^{*}(n_0)/m_N$, we should
simultaneously increase $K$ to supply positivity  of the $c$
constant in the interaction potential $U$.

The used here  values of other parameters of the SHMC model are
the same as in the KVOR one~\cite{KV04}~:

\be
z&=&0.65: \nonumber \\
C^2_\om &=& 87.600\,,\quad C_\rho^2 = 100.64\,,\quad C^2_\sigma = 179.56\,,
\nonumber\\
b &=& 7.7346 \times 10^{-3}\,,\quad c = 3.4462\times 10^{-4}\,.
\label{param_sc_065}
\ee

As we will show below, the scaling ratio $\eta_\om$ is in the
interval  $0< 1-\eta_\om \lsim 0.15$ at $T=0$ for all baryon
densities of our interest. Then $\eta_\rho$ is always finite and
positive. Considering symmetric nuclear matter $N=Z$ at a high
temperature, we will demonstrate that   $f\rightarrow 1$ (for
$T\simeq 210 $~MeV) that corresponds to $1-\eta_\om \simeq 0.32$.
However, already for a smaller value $1-\eta_\om $ the ratio
$\eta_\rho$ in (\ref{etnun1}) has the pole. Actually, in the case
$N=Z$ the $E^{\rm MF}_\rho =0$ and the problem does not arise.
Nevertheless, if we wanted to  describe the high-temperature
regime for $N\neq Z$ in a similar way as for $N= Z$, we would need
to correct the above expression for $\eta_\rho$. Thus we suggest
instead of Eq.~(\ref{etnun1}) to use its Taylor expansion \be
\label{etnun2} \eta_\rho = \eta_\om \sum_{n=0}^{10}\left[
|1-\eta_\om | (1+4\,\frac{C_\om^2}{C_\rho^2})\right]^{n}, \ee
 that solves the pole problem. To reproduce (\ref{etnun1}) in the range
$0< 1-\eta_\om \lsim 0.15$ it is sufficient to take $10$ terms in
(\ref{etnun2}). Certainly we could introduce other
parameterizations for the scaling functions. However we will use
advantages of the KVOR model have been demonstrated in
\cite{Army,KV04} in application to cold nucleon matter. Therefore
we use the choice (\ref{etnun}), (\ref{etnun2}).

Like in the KVOR model, we use a rather large value of the Dirac
effective nucleon mass at the saturation (see
Eq.~(\ref{MWnu-inp})) as compared to the most of RMF models
describing finite nuclei. The latter models usually assume $m^*_N
/m_N \simeq 0.54\div 0.7$, cf.~\cite{T05}. Only in this case these
models allow one  to appropriately describe the orbital potential
in finite nuclei. { Note that if effective meson
masses are used instead of free ones, the depth of the diffuseness
layer of the nucleus is changed, affecting the value of the
spin-orbit potential. This rises  hope that the spin-orbit
potential
problem
could be resolved if we solved space-inhomogeneous equations in
our model.
Indeed, already in the framework of the standard non-linear RMF
model  making use of a smaller value of the $\sigma$ mass, Ref.
\cite{MVTC05} permitted  to
fit the nucleon density profiles. Anyhow the description of finite
size effects as well as the solution of the mentioned problem are
beyond the scope of our consideration in the present work.
Noteworthy that one should distinguish between the Dirac and
Landau effective masses. The latter value is manifested in the
nucleon spectra. Experimental data seem to favor the Landau
effective nucleon mass being close to the free nucleon mass,
cf.~\cite{M83,Skh,RingLitv}. Recent calculations of both the Dirac
and Landau effective masses within the
Dirac-Brueckner-Hartree-Fock approach give $m_N^* /m_N \simeq 0.7$
for the Dirac mass and $\simeq 0.8$ for the Landau mass (see Fig.
2 in \cite{vanDalen}) whereas pure  RMF model calculation and that
with the energy-dependent correction produce values $0.76$ and
$0.92$, respectively \cite{RingLitv}. Since the ratio $0.92$ for
the Landau mass obtained within the model \cite{RingLitv} is still
low to explain nucleon spectra it can be considered as an argument
that within this model the corresponding ratio $m_N^* /m_N$ for
the Dirac mass should be higher than $0.76$.}

 Now  we will demonstrate that  the SHMC model describes
 the nucleon optical potential  in an optimal way. Optical potentials
 for a proton or neutron passing
through the cold ($T=0$) nuclear matter are introduced as, cf.
\cite{FL,DCM}:
\be
\label{opt} U_{\rm opt}^{N}=\epsilon
-\sqrt{{p}^2 +m_N^2},\,\,\, N=p,n~,
 \ee
where $\epsilon$ is the nucleon energy and ${p}$ is the
3-momentum. Substituting $p^2$ from equations of motion
\be
\label{optp} \sqrt{{p}^2 +m_N^{*2}}=\epsilon -g_{\om N} \
\chi_{\om} \ \om_0 -t_N^3 \ g_{\rho N} \ \chi_{\rho} \ R_0 +t_N^Q
\ V ~,
 \ee
\be
\label{optpp} U_{\rm opt}^{N}=\epsilon -\sqrt{(\epsilon
-g_{\om N} \ \chi_{\om} \ \om_0 -t_N^3 g_{\rho N} \ \chi_{\rho} \
R_0 +t_N^Q V)^2 -m_N^2 (\Phi_N^2 -1)},
 \ee
$t_N^3$ and $t_N^Q$ are given in Table 1. For the $N=Z$, $V=0$
case, proton and neutron optical potentials coincide.

\begin{figure}[thb]
\hspace*{30mm}\includegraphics[width=80mm,clip]{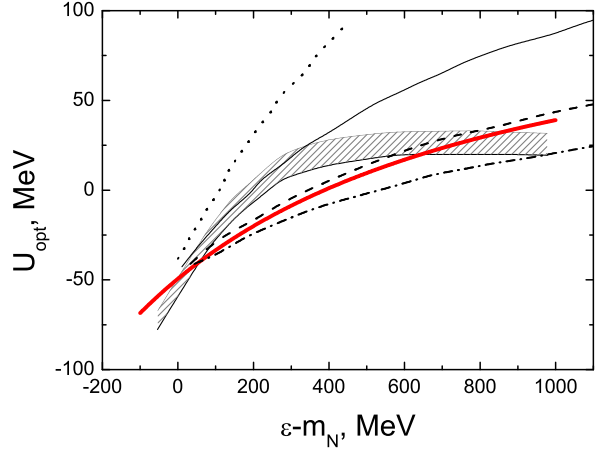}
\caption{ Energy dependence of the nucleon optical potential for
$N=Z, V=0$. Results for different values of the nucleon effective
mass (\ref{bar-m}) $m^\star_N/m_N=$0.54, 0.71, 0.8 and 0.85
calculated in the  Walecka model~\cite{FL}   are plotted by
dotted, thin continuous, dashed and dashed-dotted lines,
respectively. The solid line corresponds to our calculations with
parameter choice (\ref{MWnu-inp}). Shaded area shows uncertainties
in extrapolation from finite nuclei to cold nuclear
matter~\cite{FL}. }
 \label{Uopt}
\end{figure}

 Energy dependence of the nucleon optical
potential for $n_B=n_0$, $N=Z$ is shown in Fig.~\ref{Uopt}.
Calculated results are presented at $T=0$ and therefore coincide
with those for the KVOR model. The band is the optical potential
extracted from the data ~\cite{H90} and recalculated to the case
of the infinite nuclear $N=Z$ matter in~\cite{FL}. Different lines
are RMF calculations within the standard Walecka model for various
effective nucleon masses $m^*_N /m_N =0.54, 0.71, 0.8,
0.85$~\cite{FL}. We see that the resulting optical potential of
the KVOR model is the closest to that with $m^\star_N/m_N=$0.8
of~\cite{FL}. Since in application to HIC particle spectra should
be  described in a large energy range (up to and above proton
momenta $\sim$ 1 \ GeV$/c$) in average the KVOR description of the
data is optimal.

The isovector
part of the optical potential $U_{\rm opt}^{n}-U_{\rm opt}^{p}$ is less
constrained by the data, cf. \cite{Fuchs}. Therefore we do not consider
it here.

The   ratios $x_{mb}$ (see Eq.~(\ref{rat})) are not well fixed
experimentally. Different possible choices are reviewed in
\cite{Gphysrep,PREPL}. Within a quark counting model (cf. "case I"
of Ref. \cite{KV03}) one gets $x_{mb}=1$ for non-strange baryons
(in our case $\Delta$'s) and
 \be
 \label{rat-count}
x_{\omega \Lambda} = x_{\omega \Sigma} = x_{\omega \Sigma^{*}} =
x_{\rho \Sigma}=x_{\rho \Sigma^{*}}= x_{\rho \Lambda}= 2x_{\omega
\Xi}=2x_{\rho\Xi}=\frac{2}{3}
\ee
in the strange sector, whereas
the constituent $SU(6)$ quark model, gives different values, e.g.,
$x_{\rho \Xi}=1$, $x_{\rho \Lambda}=0$. In this paper we use the
quark counting values first and then allow for a variation of
these parameters.

The ratios $x_{\sigma h}$  are determined with the help of the
relations (compare with~\cite{KV03}):
\be
E^h_{\rm bind}(n_0) =
(g_{\omega N}^2 n_0 /[\eta_{\om} (n_0) \ m_{\omega}^2 ]) x_{\omega
h}-(m_N -m_N^{*}(n_0)) \ x_{\sigma h} ,
\ee
where $E^h_{\rm
bind}(n_0)$ is the binding energy for the hyperon $h$. There is a
convincing evidence from the systematic study of hypernuclei that
for $\Lambda$ particles $E_{\rm bind}^\Lambda \simeq -30$~MeV. For
$\Xi$ we adopt the value $E_{\rm bind}^{\Xi}\simeq -18$~MeV and
for $\Sigma$ hyperon $E_{\rm bind}^{\Sigma} \simeq -10$~MeV
(following "case I" of~\cite{KV03}) is taken. There are no
experimental data for $\Sigma^{*}$. For $\Sigma^{*}$ we take the
same value as for $\Sigma$. One could use other parameter choices
within experimental error bars, e.g. following "cases II-IV" of
Ref.~\cite{KV03}.

We put the coupling $g_{\om\pi}=0$, because $\om$ does not decay
in two pions. As follows from the  $\pi^-$ atomic data, one needs
a slight $\pi^-$ energy  shift  upward, $\sim 10\div 30$~MeV for
$N=Z$ at the saturation density $n_B=n_0$, cf. \cite{KKW}.  The
value of the pion $\Sigma$-term estimated from   scattering data
is rather small: $\Sigma_{\pi N}\simeq 30\div 45$~MeV. Since both
these values (energy shift and $\Sigma_{\pi N}$) are small,  we
can assume $g_{\om\pi}=g_{\sigma\pi}\simeq 0$. The $p$-wave $\pi
NN$ and $\pi N\Delta$ interactions are disregarded within our
RMF-based model, as all other $p$-wave effects. Thus for $N=Z$ we
deal with free pions.

The $\rho\pi$-meson coupling  $g_{\rho\pi}$, which is necessary to
describe the pion behavior in isotopically asymmetric matter, can
be found by matching with the $s$-wave Weinberg-Tomazawa term of
$\pi^-$ polarization operator
\be
g_{\rho\pi} R_0 \rightarrow
n_n /(2f_{\pi}^2) \quad \mbox{for}  \quad n_B\rightarrow 0, \ Z=0,
 \ee
as it is motivated by the chiral symmetry, $f_{\pi}=93$~MeV, cf.
\cite{MSTV90}. Thus one gets
\be g_{\rho\pi}\simeq
\frac{m_{\rho}}{2g_{\rho N} \ f_{\pi}^2}~.
\ee
The corresponding
value $g_{\rho\pi}\simeq 6$ is consistent with that
follows from the universality relation, cf. \cite{bando83}.

For the kaon coupling constants, we take $g_{\omega K}=g_{\omega
N}/3$, $g_{\rho K}=\frac{1}{2}g_{\rho N}$ as follows from quark
counting, cf. \cite{Gphysrep}, and   $g_{\sigma K}$ is evaluated
from the $K^+$ and $K^-$  optical potentials at $n_B =n_0$ :
$$U_{\rm opt}^{(K^+ /K^- )}(n_0 ,T=0
)=-g_{\sigma K} \ \sigma (n_0 ,T=0)\pm g_{\omega K} \ \omega_0
(n_0 ,T=0)~.$$ The experimental values of $U_{\rm opt}^{K^+}(n_0
,T=0 )$ are in the range $\simeq 20\div 30$~MeV.
 It is known that the $\Sigma_{KN}$ sigma term is significantly
 larger than $\Sigma_{\pi N}$. However the former quantity is not
 well determined and may vary in a broad range, $\Sigma_{KN}\sim
 150\div 400~$MeV. The values of a deep potential
$U_{\rm opt}^{K^-}(n_0 ,T=0 )\simeq -100\div -200$~MeV are derived
in Refs.~\cite{FG99} from analysis of kaonic atom phenomenology,
whereas self-consistent calculations based on a chiral
Lagrangian~\cite{LK02} and coupled-channel $G$ matrix theory
within meson exchange potentials~\cite{TRP} yield $U_{\rm
opt}^{K^-}(n_0 ,T=0 )\simeq -50\div -80$~MeV. In Ref.~\cite{KV03}
the effective value $\Sigma_{KN} \simeq 150$~MeV is extracted
using the analysis of the kaon-nucleon scattering of \cite{LK02}.
Thereby, we apply a shallow $K^-$ potential  $U_{\rm
opt}^{K^-}(n_0 ,T=0)\simeq -80$ MeV. For the $K^+$ potential we
take $U_{\rm opt}^{K^+}(n_0 ,T=0 )= 21.4$ MeV. With these
potentials we find coupling constants
 \be \label{A} (A):
\mbox{\hspace*{10mm}}g_{\sigma K}^*=0.13 \ g_{\sigma
N},\mbox{\hspace*{10mm}} g_{\om K}^*=\frac13 g_{\om N}.
\ee
 The density dependence of the $U_{\rm opt}^{K^-}$ and $U_{\rm
opt}^{K^+}$ is even less constrained. Thus we also use an another
set of kaon couplings obtained with the help of the scaling
\be
\label{B} (B):\mbox{\hspace*{5mm}}g_{\sigma K}^*=0.13 \ g_{\sigma
N}\ \chi_\sigma/\chi_\sigma (n_0), \mbox{\hspace*{10mm}} g_{\om
K}^*=\frac13 g_{\om N}\ \chi_\om/\chi_\om (n_0)
\ee
in accordance
with the above scaling hypothesis.

\begin{figure}[thb]
\hspace*{30mm}\includegraphics[width=80mm,clip]{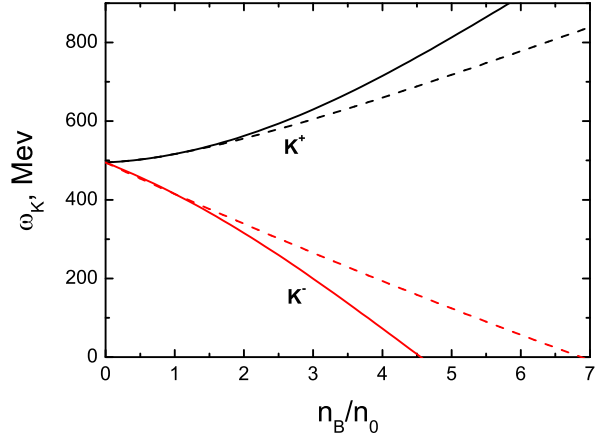} \caption{
Kaon and antikaon energies for ${\vec p}=0$ as a function of the
baryon density for $N=Z$, $T=0$, $V=0$ for two sets of couplings:
Solid curves are for the $(A)$ set without scaling, Eq.~(\ref{A}),
and dashed curves are for the $(B)$ set with scaling,
Eq.~(\ref{B}). }
 \label{omKocmp}
\end{figure}

In Fig.~\ref{omKocmp}  the kaon and antikaon dispersion curves are
shown versus  the baryon density for iso-symmetric system at
vanishing temperature for both  sets of couplings.
 The difference between the kaon (and antikaon) dispersion curves
 for these two parameter choices is tiny up to $2n_0$ and becomes
 significant at higher baryon densities.
Energies of antikaons $\om_{K^-}$ reach zero at $n_B\simeq 4.5n_0$
and $\simeq 7n_0$ for   parameter sets $(A)$ and $(B)$,
respectively. If a deeper $K^-$ optical potential is used, as
suggested in~\cite{FG99}, one obtains that $\om_{K^-}=0$ at
smaller density.  We note that the points $\om_{K^-}=0$ are not
the critical points of a antikaon condensation. In HIC strangeness
is conserved. Then  $K^-$ mesons can be created only in pairs with
$K^+$ mesons at $T=0$, if the hyperon Fermi seas are not filled.
Then  the condensation condition for kaons and antikaons looks
like $\om_{K^{\pm}} (p=0) \pm \mu_{\rm str}=0$. In the case of
infinitely long-living matter the strangeness is not conserved,
and  we would deal with the kaon condensation at $\om_{K^-} =0$
for $T=0$. The consideration of the dense $N=Z$ system at $T=0$
has only pedagogical interest, however.

\begin{figure}[thb]
\hspace*{30mm}\includegraphics[width=90mm,clip]{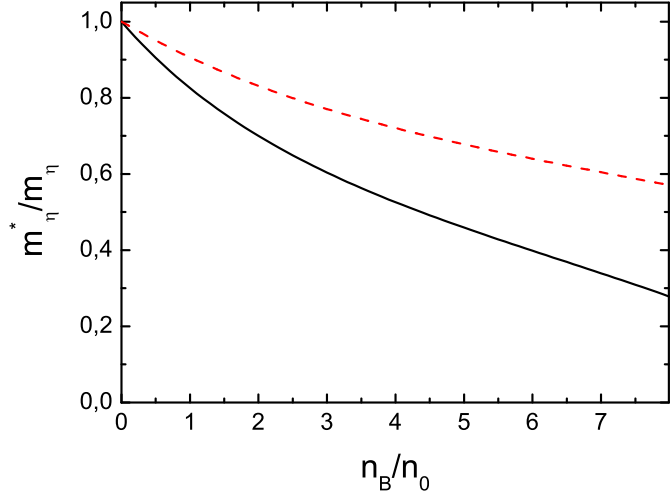}
\caption{ Effective $\eta$ meson mass given by Eq. (\ref{etam}) in Appendix A
 as a function of the baryon
density for $N=Z$, $T=0$, for two sets of couplings. Solid curve:
$\Sigma_{\eta N}=280~$MeV, $\kappa_{\eta N} =0.4~$fm. Dash curve:
$\Sigma_{\eta N}=140~$MeV, $\kappa_{\eta N} =0.2~$fm.}
 \label{Meta}
\end{figure}

Effect of the $\eta$-meson on thermodynamic quantities is minor.
 In our calculations of characteristics of HIC
 we  take  $\Sigma_{\eta N}\simeq 140$~MeV,
$\kappa_{\eta N} =0.2$~fm (see Eq. (\ref{Leta})).
These values are  twice smaller than the
average-weighted values  $\Sigma_{\eta N}\simeq 280$~MeV,
$\kappa_{\eta N} =0.4$~fm
 used in Ref. \cite{ZPLN}. We make this choice  in order to
simplify the consideration by avoiding a possibility of $\eta$
condensation in a wide baryon density - temperature range of our
interest. Differences of these two parametrization in the $\eta$
meson mass are shown in Fig.~\ref{Meta}.

In case when there is no  information on
 interactions of a particle with mean $\sigma$-, $\omega$- and
$\rho$-meson fields, or if it is known that this interaction is
rather weak, we treat this particle as a free one. As we have
mentioned, we consider $\eta^{\prime} (958)$, $K^* (892)$ and
$\varphi (1020)$ as free particles. We also consider $\Omega$ and
$\Xi^*$ as free ones. For $\Sigma^*$ we use the same couplings as
for $\Sigma$.

Though  the  constructed model describes arbitrary iso-asymmetric
systems, in this paper we further focus on the study of
isospin-symmetric nuclear matter, $N=Z$, and disregard small
Coulomb effects. In calculations presented below we use by default
parameters (\ref{param_sc_065}), (\ref{rat-count}), (\ref{B}),
unless other is specified.

\section{SHMC EoS  for $T=0$ and $N=Z$}\label{EoS0}

At vanishing temperature our model differs from that of KVOR
\cite{KV04} in two aspects: (i) we take into account a possibility
of occupation of the Fermi seas by different baryon species at
higher baryon densities; (ii) we incorporate a possibility of
condensation of the (quasi)Goldstone boson fields, when it occurs.

\begin{figure}[thb]
\hspace*{-2mm}\includegraphics[width=70mm,clip]{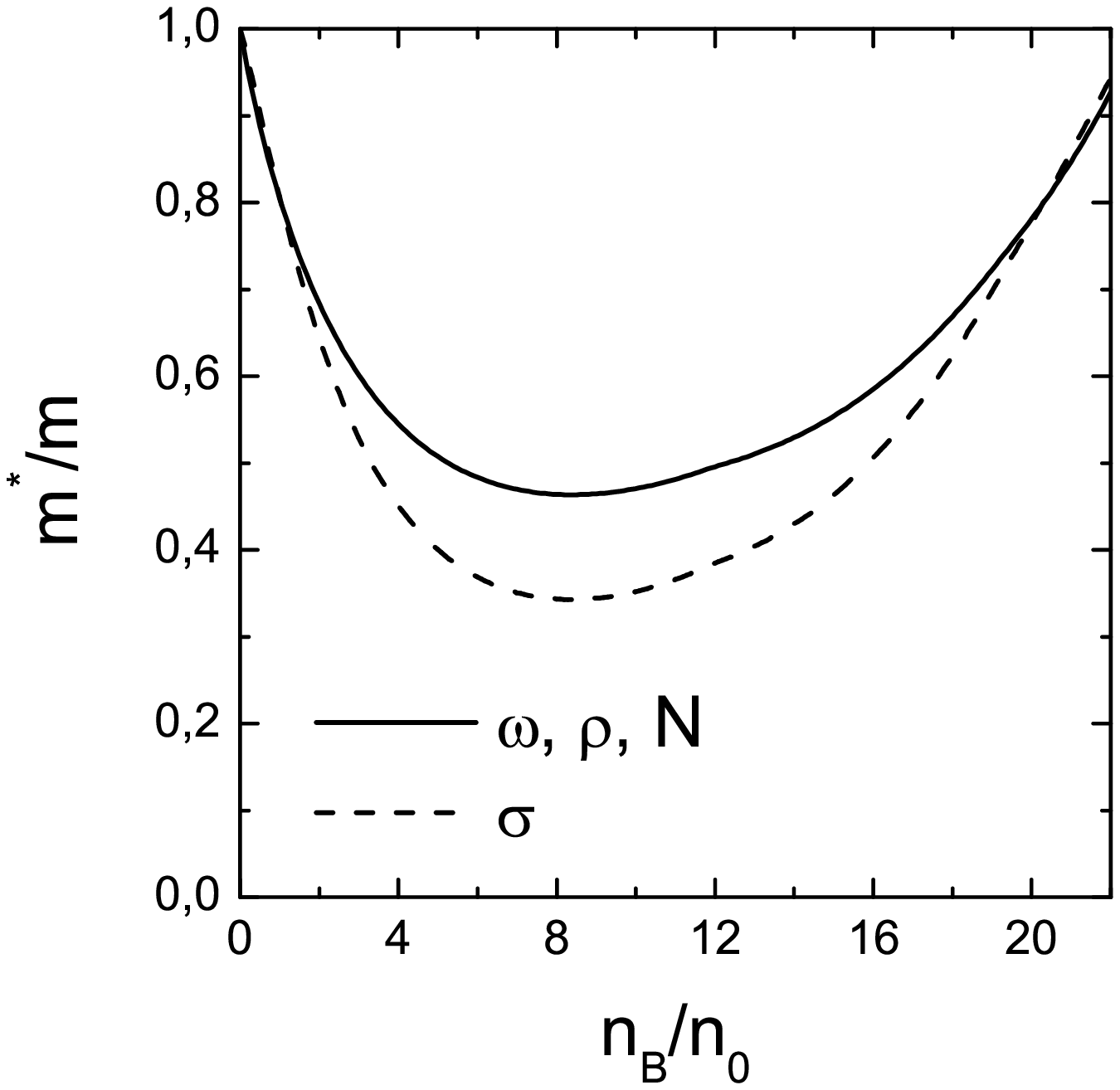}
\includegraphics[width=70mm,clip]{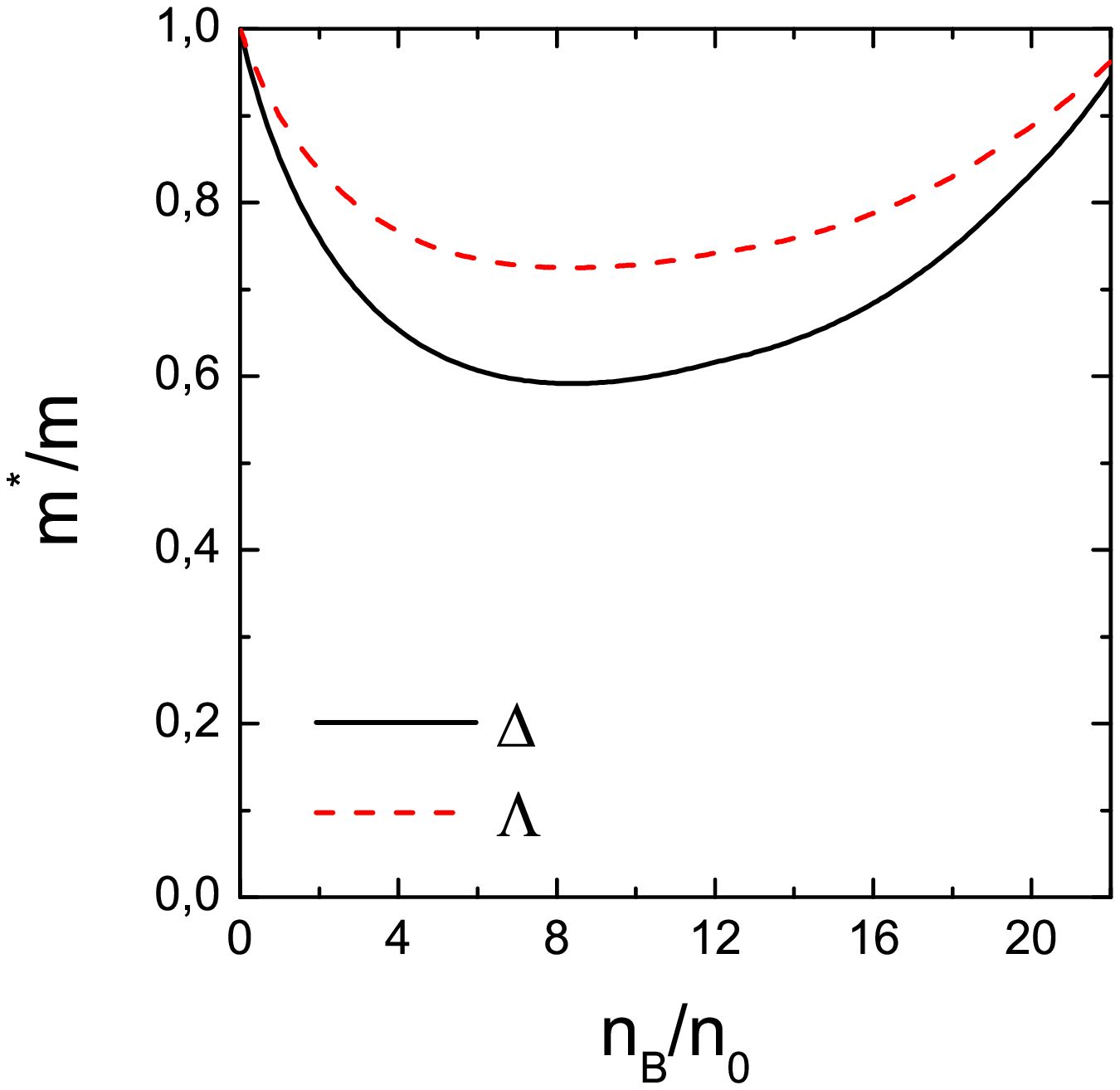}
\caption{ Baryon density dependence  of the effective-to-bare mass
ratio at $T=0$, $N=Z$. Left panel: for nucleon-
$\om$-$\rho$ excitations (solid line),  and for $\sigma$
excitations (dashed line). Right panel: The same as in left panel
but for the $\Delta (1232)$ isobar (solid line) and  the $\Lambda
(1116)$ hyperon (dashed line). }
 \label{Mz0}
\end{figure}

The left panel of Fig.~\ref{Mz0} shows the baryon density
dependence of the ratio of the effective mass to the bare mass for
nucleons and  $\om$ and $\rho$ excitations, $m_N^* /m_N
=m_\om^{\rm part *} /m_\om =m_\rho^{\rm part *} /m_\rho =\Phi$ for
$\Phi >0$, see Eqs.~(\ref{bar-m}) - (\ref{Br-sc}), and
(\ref{omr}), as well as for $\sigma$  excitations,
$m_{\sigma}^{\rm part *} /m_{\sigma}$,
 given by Eq. (\ref{spa}) for $T=0$.

We observe that the effective masses monotonically decrease up to
a minimal value at the density $n_B <n_{min,B}\simeq 8 n_0$ and
then begin to grow. This is a consequence of the fact that within
our model the masses depend non-linearly on the $\sigma$ field and
this dependence is determined within self-consistent calculations.
Due to this feature the SHMC model EoS is getting stiffer with
increasing baryon density in the range $n_B < n_{min,B}$ and then
it becomes softer for $n_B > n_{min,B}$. Such a high-density
behavior could additionally favor the deconfinment phase
transition at large densities ($n_B > n_{min,B}$) at $T=0$ if it
had not yet happened at a smaller density. We could chose model
parameters in such a way that the value of $n_{min,B}$ would be
smaller, in favor of shifting the position of the deconfinment
transition to smaller density. However it would result in the
simultaneous decrease of the maximum neutron star mass
$<1.9~M_{\odot}$. The latter may come in conflict with
experimental data on neutron stars, see \cite{Army}.

Note that the non-linear
density dependence of the nucleon and $\sigma$ effective masses
resembles the density dependence of the chiral and gluon condensates
obtained in Refs.
 \cite{BW,RK}.

 In the right panel of Fig.~\ref{Mz0} we show
effective masses of the $\Delta (1232)$ isobar and the $\Lambda
(1116)$ hyperon, as  representative examples of heavy baryon
species. Their baryon density dependence is similar to that for
the nucleon,  with  the same value of $n_{min,B}$. However the
value of $m^*_b (n_{min,B})/m_b$ ($b\ne N$) is higher than $m^*_N
(n_{min,B})/m_N$.

\begin{figure}
\hspace*{30mm}\includegraphics[width=80mm,clip]{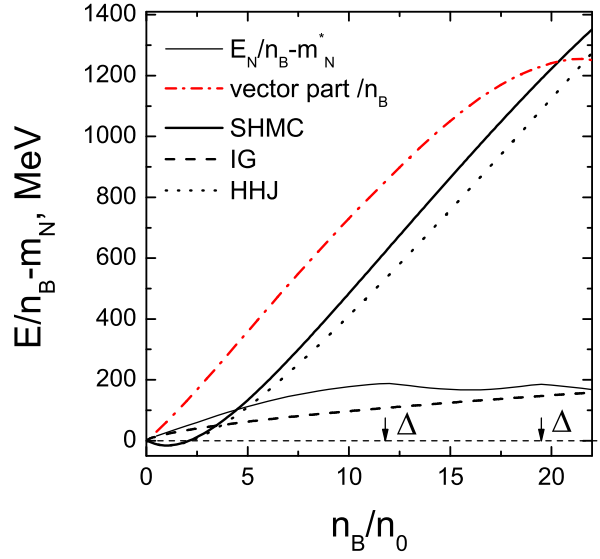}
\caption{Baryon  density dependence of the total energy per baryon
(solid line)  for SHMC EoS, for A18+$\delta v$+UIX* in the HHJ
parametrization \cite{HHJ} (dotted line), and  for the IG EoS
(dashed line); $T=0$, $N=Z$. Arrows show the threshold densities
for the appearance and melting of the $\Delta$ isobar Fermi sea.
Dashed-dotted curve is a partial contribution of the repulsive
vector field term of the SHMC EoS. For comparison the kinetic
energy $E_N /n_B-m^\star_N$ for the SHMC model is shown by the
thin continuous line.
 }
 \label{Ebind0}
\end{figure}

\begin{figure}
\hspace*{30mm}\includegraphics[width=80mm,clip]{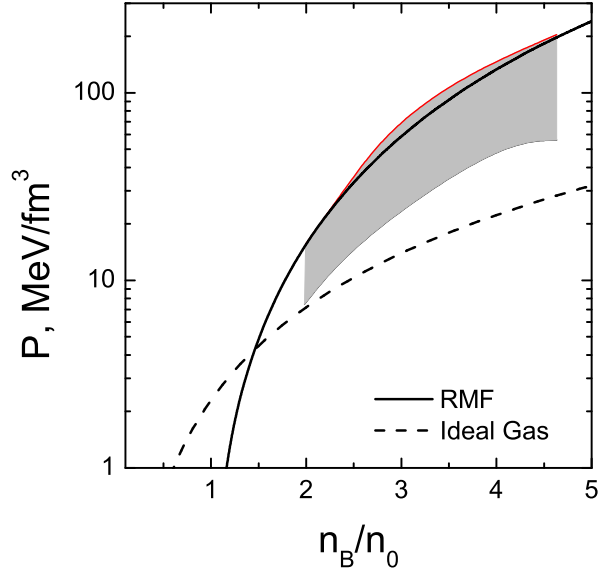}
\caption{Baryon density dependence of pressure for symmetric
nuclear matter at $T=0$ for the SHMC (solid line) and  IG (dashed)
EoS. The shaded area corresponds to the experimental constraint
derived from HIC~\cite{Dan02}.
 }
 \label{p0}
\end{figure}

In Fig.~\ref {Ebind0} we show the density-dependent  total energy
per baryon for SHMC EoS  in comparison with that for the
Urbana-Argone (A18+$\delta v$+UIX*) EoS \cite{APR98} (in the HHJ
version of Ref.~\cite{HHJ})  and for the ideal gas (IG) EoS.
Within the IG model we include the same particle species, as in
the SHMC model, but in this case all mean fields and thus all
particle interactions are switched off. Thus, in the IG model at
$T=0$ only nucleon  Fermi seas contribute. It is seen that
difference between  SHMC EoS and IG EoS grows strongly  with the
density increase   indicating to an important contribution of
particle interactions. The threshold density for the appearance of
the $\Delta (1232)$ isobars is $\sim 12~n_0$ and their Fermi sea
again melts for $n_B >20~n_0$ (these densities are shown by arrows
in Fig.~\ref{Ebind0}). Appearance of $\Delta$'s almost does not
affect the total energy per baryon. Hyperons do not occur at all
for $N=Z$ in contrast with $\beta$-equilibrium matter,
cf.~\cite{KV03}.

 Though baryon and meson masses begin to
increase for $n_B >n_{min ,B}$, this only moderately affects the
stiffness of the EoS, since the suppression of the nucleon kinetic
term (thin solid  line in Fig. \ref{Ebind0}) is largely
compensated by the increase of the repulsive vector meson term
(dashed-dotted line in Fig. \ref{Ebind0}).  Due to that SHMC EoS
remains stiffer for $n>n_{min ,B}$ compared to the HHJ EoS (dotted
line in
 Fig. \ref{Ebind0}).

The SHMC EoS begins to differ from the HHJ EoS for $n_B >4n_0$ and
this difference increases with increase of the baryon density.
Such a behavior, cf. \cite{KV04}, results in   an increase of the
value of the maximum neutron star mass, $M_{max} =2M_{\odot}$,
that is in agreement with the value $M_{max} =(2.1\pm
0.2)M_{\odot}$ (at the $1\sigma$ confidence level)  derived in
\cite{NiSp05} from the observations of the PSR J0751+1807, a
millisecond pulsar in a binary system with a helium white dwarf
secondary.

In Fig.~\ref{p0} the  pressure calculated in the SHMC model (solid
line) is compared with the experimental constraints coming  from
the analysis of elliptic flow in HIC~\cite{Dan02}. As it has been
argued in \cite{Army}, only the EoS with pressure curves, being
close to the upper boundary of the band, satisfies the maximum
neutron star mass constraint.  Pressure within the IG model of EoS
(dashed line) does not fulfill the HIC flow constraint. To satisfy
the flow constraint at $T=0$, one definitely needs a much stiffer
EoS  than that given by the IG model.

{ Transport calculations \cite{F} have demonstrated
that subthreshold $K^+$ production may provide an important
information to constrain the EoS of the warm symmetric nuclear
matter for $n_B \lsim 3~n_0$. Within the last decade the KAOS
collaboration at GSI performed measurements of the kaon production
\cite{Kaos}. Analysis of the data \cite{F1} led to a conclusion
that the  EoS satisfying the  kaon data is compatible with the
 above-required  flow constraint. Both constraints hold true with the EoS
of the Urbana-Argonne group (${\rm{ A18}}+\delta v+
{\rm{UIX}}^{*}$) and with the KVOR-based SHMC EoS used here.}

\section{SHMC EoS  for $T\neq 0$ and $N=Z$}\label{EoST}

\subsection{Density-temperature dependence of effective masses of
 excitations}

\begin{figure}[thb]
\hspace{-2mm}\includegraphics[width=70mm,clip]{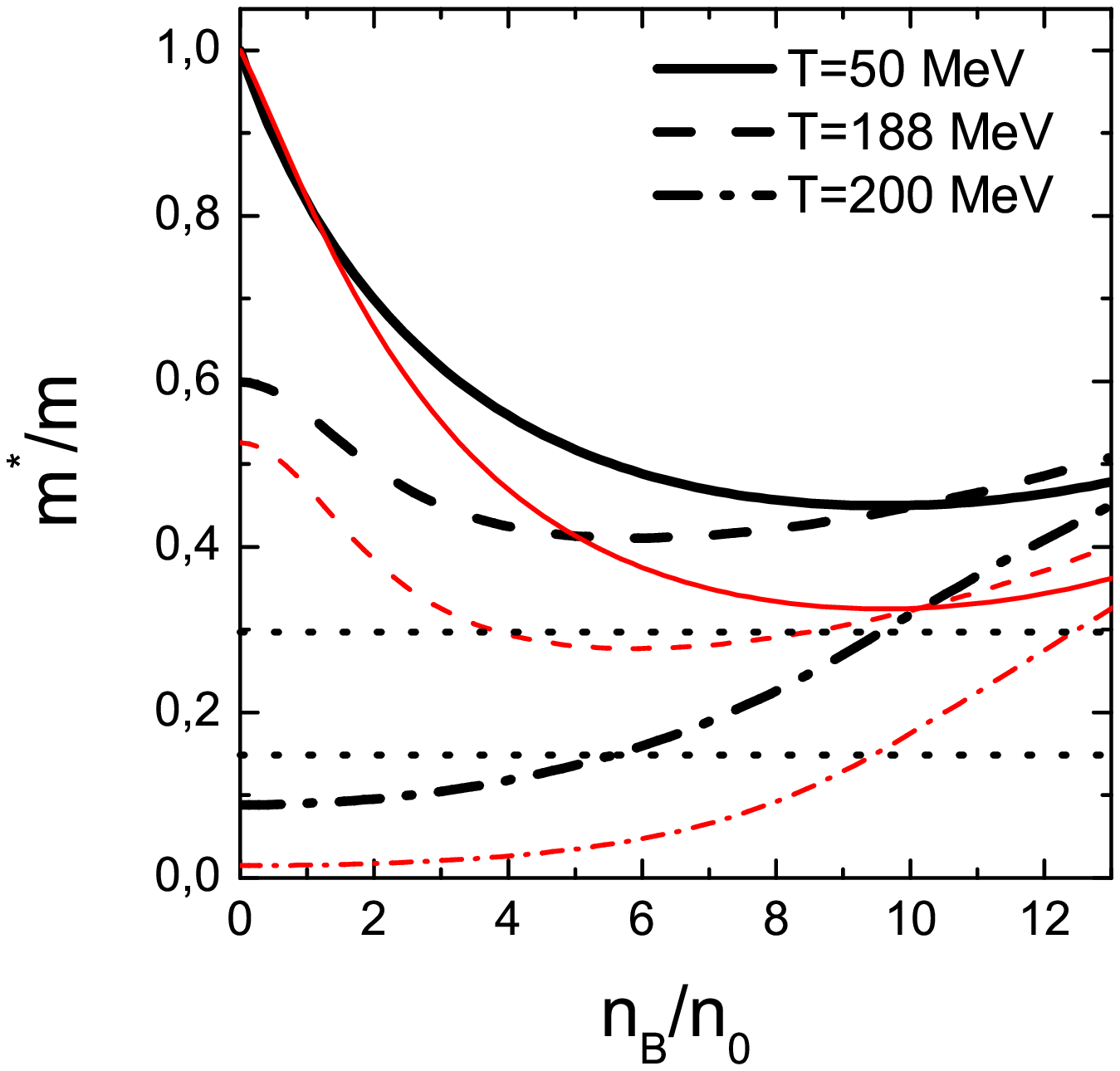}
\includegraphics[width=70mm,clip]{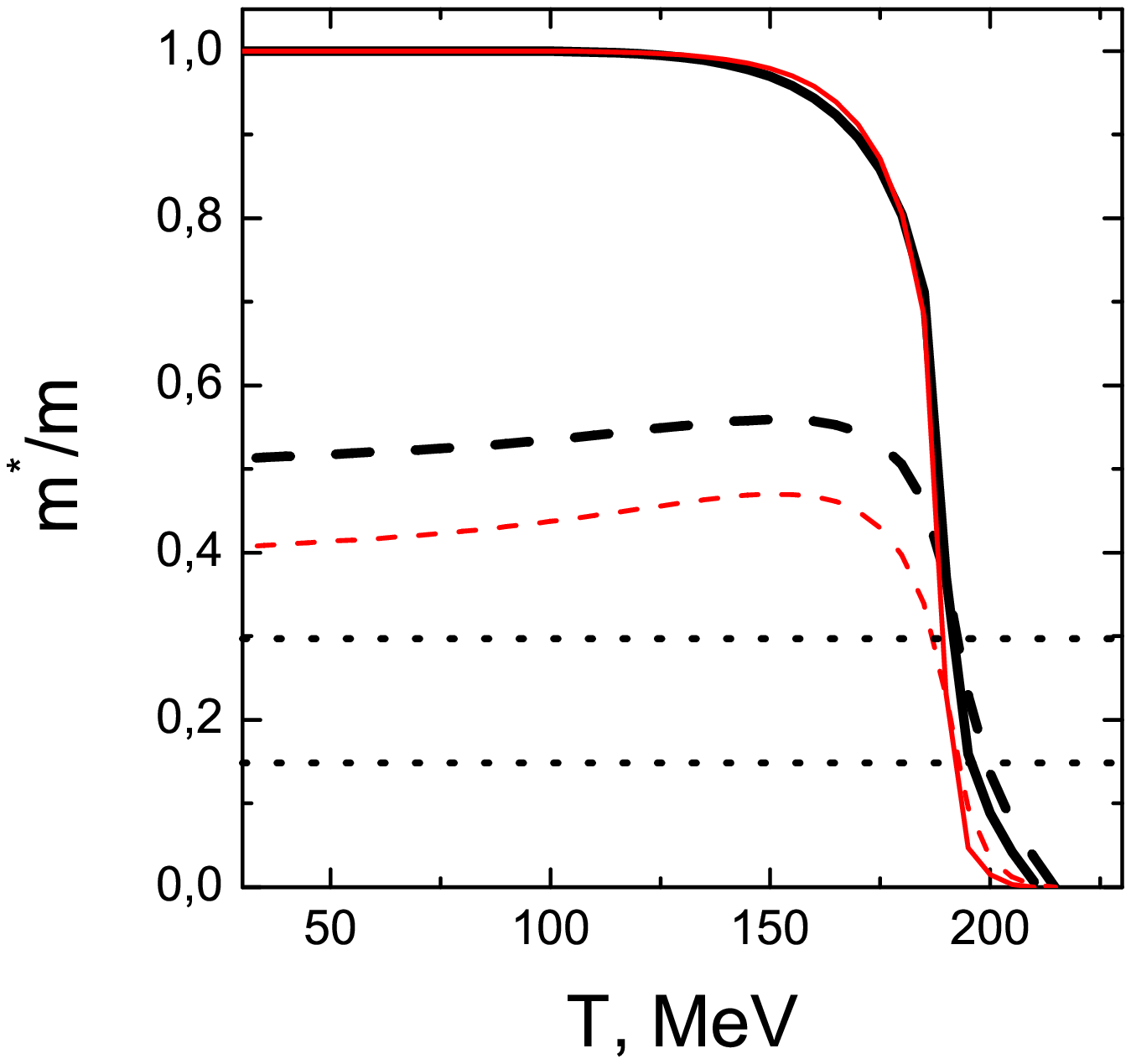}
\caption{ The  effective masses of the nucleon,
 $\om$, $\rho$ excitations (thick lines), and of
$\sigma$ meson  excitations (thin lines), $N=Z$. Two dotted straight
lines show the $\sigma$ mass levels $m_\sigma=2m_\pi$ and
$m_\sigma=m_\pi$ (bottom). Left panel: The baryon density
dependence at different temperatures.  Right panel: Temperature
dependence at the baryon density $n_B=0$ (continuous lines) and
$n_B=5n_0$ (dashed lines). }
 \label{Mz}
\end{figure}

The effective masses of the nucleon and $\omega$/$\rho$
excitations follow the same scaling law and coincide, see
Eqs.~(\ref{Br-sc}), (\ref{spa}), (\ref{omr}). As it is seen from
the left panel of Fig.\ref{Mz}, at the baryon density $n_B <6\div
8~n_0$ and $T\lsim 190$~MeV, the effective masses of the nucleon
and $\omega$/$\rho$ excitations and the $\sigma$ excitation
decrease when the density grows up and then they start to increase
at higher densities similarly to the case $T=0$.

Generally, different phase states can be realized within the SHMC
model.  At some density $n_B =n_B^{\sigma\pi}$ and  temperature
$T=T^{\sigma\pi}$ the $\sigma$ excitation mass  may reach the
value $2m_{\pi}$. Then the decay $\sigma \rightarrow 2\pi$ becomes
forbidden at higher $T$ and $\mu_B$. Ref.~\cite{S05} argued that
due to  long-scale field fluctuations the scattering length of two
pions at rest should go to infinity at $n_B\rightarrow
n_B^{\sigma\pi}$,  identically to the so called "Feshbach
resonance" at zero energy to be used in atomic physics for cold
trapped atoms. The result is also known as a new "strong coupling"
regime of matter  which manifests a liquid-like behavior
\cite{Hara}. However these interesting questions are beyond the
scope of the present paper since in the SHMC model the particle
excitations have no widths. Therefore  we continue to apply our
model without any modifications also for $n_B >n_B^{\sigma\pi}$.

In addition, the $\sigma$ excitation mass  reaches the value
$m_{\pi}$ at some density $n_B^{\rm chir}$ and temperature
$T=T^{\rm chir}$.  Note that in models with inherent chiral
symmetry the $\sigma$ and $\pi$ masses meet at the chiral symmetry
restoration point. As it is seen from  Fig.~\ref{Mz}, with the
SHMC model for the $T=0$ case the $\sigma$ excitation mass remains
always higher than the double pion mass. But  at sufficiently high
temperature, $T\gsim 190~$ MeV, both points, $n_B
=n_B^{\sigma\pi}$ and $n_B^{\rm chir}$, are reached.

As follows from the left panel of Fig.~\ref{Mz}, for the phase
range $T\lsim 190~$MeV and $n_B <(6\div 8)~n_0$ the density
behavior of effective masses is similar to that in models with
inherent partial restoration of the chiral symmetry. The values of
effective masses decrease at low densities with the temperature
increase in accordance with behavior of $\Phi=1-f$. For $T>
190$~MeV all the effective masses are getting small rather sharply
but then grow slowly with the increasing density. At temperature
$T= $188 MeV the critical value $n_B^{\sigma\pi}$ is reached first
at $n_B^{\sigma\pi}\approx 4n_0$ and this phase state is left at
$n_B^{\sigma\pi}\approx 8n_B$ (see crossing of the thin dashed
line with the horizontal one). At $T=200~$MeV the mass of the
$\sigma$ field is less than $m_\pi$ for $0\leq n_B \lsim 10n_0$.

As it is seen from the right panel of Fig. \ref{Mz}, the
temperature dependence of the effective masses of nucleon and
sigma excitations   is small up to $T\sim 170\div 180$ MeV. For
higher temperatures the effective masses begin to decrease
abruptly and for $n_B=0$ we have $T^{\sigma\pi} \approx T^{\rm
chir}\approx 190~$MeV. If one proceeds to the dense matter
($n_B=5n_0$) the difference between these temperatures is about
few MeV. Within this narrow temperature interval the second
derivative of the effective mass with respect to temperature
changes the sign. The effective nucleon-$\om$-$\rho$ and $\sigma$
excitation masses reach zero at the same critical temperature
$T_{c\sigma}$, about 210 MeV. Since the coupling scaling functions
$\chi_{\sigma}$ and $\chi_{\om}$ follow the same dropping trend as
the mass scaling function $\Phi$, in vicinity of $T_{c\sigma}$ we
deal with a gas of almost massless excitations. Similar result has
been obtained in \cite{BGR} using a generalized local symmetry
approach and vector manifestation arguments. The anti-nucleon
yield rapidly increases due to a sharp decrease of the nucleon
mass.

For $T>T_{c\sigma}$ the effective nucleon mass becomes negative.
For the first time such a behavior of the nucleon effective mass
has been found within the standard RMF model including $\Delta$
resonance in~\cite{WTMSG}. Authors suggested a specific choice for
the resonance-$\sigma$ couplings that allows to restore
positiveness of masses. Actually in the region where the effective
nucleon mass is negative nothing dramatic happens. The nucleon
spectrum given by Eqs.~(\ref{Eb}), (\ref{oc}) continues to be well
defined since these equations enters $m_N^{*2}$ rather than
$m_N^{*}$.

Note that in our model the effective $\om$ and $\rho$ excitation
masses follow the law (\ref{Br-sc}), (\ref{omr}). Thus they only
touch zero at $T=T_{c\sigma}$ and become  again positive for
$T>T_{c\sigma}$. Therefore $\om$ and $\rho$  condensates should
not appear at $T>T_{c\sigma}$.

However the effective mass of the $\sigma$ excitation
($m_{\sigma}^{\rm part*}$, see Eq. (\ref{sigm-mass} ) in Appendix
A) is getting imaginary at $T>T_{c\sigma}$. Thus, the ground state
proves to be unstable  with respect to the Bose condensation of
the $\sigma$ excitation field. The stability is achieved due to
the self-interaction between the $\sigma$-particle excitations,
see Appendix B. Such a condensation might be called "hot Bose
condensation" since it occurs at $T>T_{c\sigma}$, in contrast with
the standard Bose-Einstein condensation appearing with the
temperature decrease. A possibility of hot Bose condensation has
been considered in~\cite{V04}, within a different model which
includes effects of particle widths. In order not to complicate
consideration we avoid description of temperature region above
$T_{c\sigma}$ in the present work.

\begin{figure}[thb]
 \hspace*{2mm}
\includegraphics[width=130mm,clip]{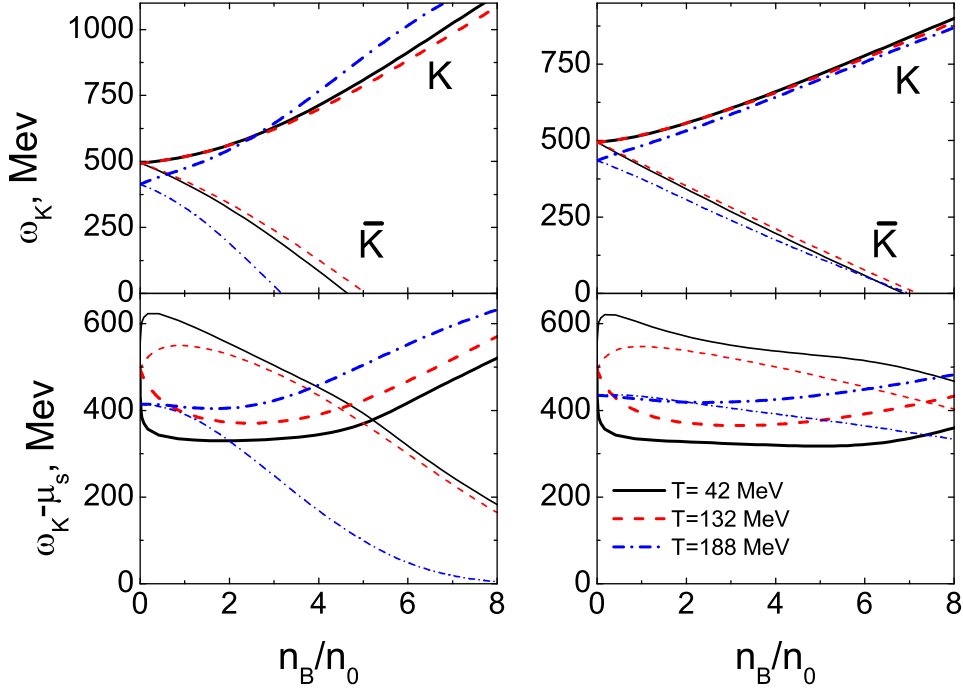}
\caption{ Baryon density dependence of energies $\om_K$ (upper
panels)) and $\om_K-\mu_s$ (lower panels) for  kaons (thick lines)
and antikaons (thin lines) at ${\vec p}=0$ at three values of the
temperature $T=$42, 132 and 188~MeV, $N=Z$.  Left panel:
 the $(A)$ set for kaon couplings without scaling,
Eq.~(\ref{A}). Right panel: for the $(B)$ set of couplings with
scaling, Eq.~(\ref{B}).
 }
 \label{omK}
\end{figure}

Fig.~\ref{omK} (upper panels) presents the baryon  density
dependence of the energy for kaons ($K^+$ or $K^0$) and antikaons
($K^-$ or $\bar{K}^0$) with zero momenta at three values of
temperature, see Eqs.~(\ref{EK+})-(\ref{EK0}) of Appendix A.
 Kaons ($K^+$ and ${K}^0$)  (as well as antikaons ($K^-$ and
$\bar{K}^0$)) have the same dispersion relations in the
iso-symmetrical matter, if  a small Coulomb contribution is
neglected. We see that the $K^+$ energy  only slightly depends on
the temperature for $T\lsim 170~$MeV, $\om_{K^+}$ decreases with
the $T$ increase. Temperature dependence of the $K^-$ energy is
also minor for $T\lsim 170~$MeV, $\om_{K^-}$ increases with the
$T$ increase. Thus for $T\lsim 170~$MeV the density dependence
remains similar to that at $T=0$, see Fig.~\ref{omKocmp}. For
higher temperatures the $T$ dependence becomes significant for
both kaons and antikaons. At $n_B =0$, branches $\om_{K^+}$ and
$\om_{K^-}$ coincide. For set $B$ of scaled couplings the
dispersion curves are more flat than for  set $A$. For set $A$,
$\om_{K^-}$ vanishes at $n_B\sim (3\div 5) n_0$ (for $T \leq
188~$MeV) and for set $B$, for $n_B\sim 7n_0$. However it does not
mean that the condensation occurs. The necessary condition for
condensation is $\om_{K}({\vec p}=0)-\mu_s=0$, $\mu_s=\mu_{\rm
str}$ for kaons and $\mu_s=-\mu_{\rm str}$ for antikaons, see
Eq.~(\ref{Tmubos}) of Appendix A. This difference is plotted in
the bottom part of Fig.~\ref{omK}. It is seen that the kaon
condensation condition is never fulfilled. Antikaon condensation
takes place only if the density-dependent scaling is neglected
(set $A$), at $n_B\sim 8n_0$ and $T\sim 190~$MeV. With set $B$ of
couplings the antikaon condensation does not occur in the relevant
density-temperature range. Therefore we perform subsequent
calculations of HIC for the case $B$.

\begin{figure}[thb]
\hspace*{30mm}\includegraphics[width=80mm,clip]{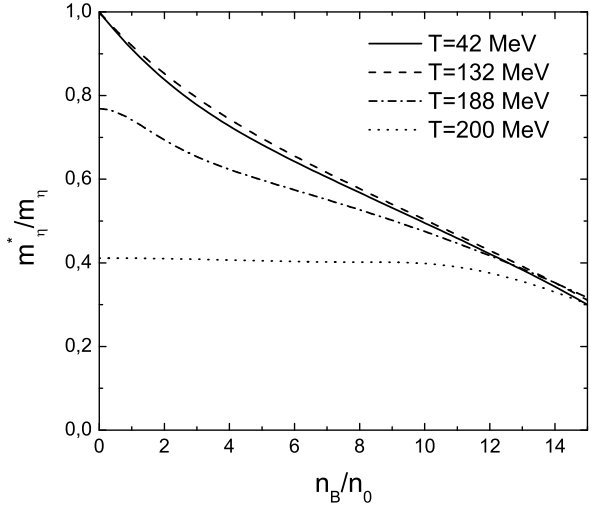}
\caption{ Baryon density dependence of the $\eta$  effective mass
at four values of the temperature $T=$42, 132, 188 and 200~MeV,
$N=Z$, for $\Sigma_{\eta N}=140$~MeV and $\kappa_{\eta N}=0.2$~fm.
}
 \label{MetaT200}
\end{figure}

Fig.~\ref{MetaT200}  shows that the $\eta$ effective mass
monotonically falls down with the baryon density if the
temperature is not very high. However at $T\sim 200~$MeV the
 $\eta$ effective mass becomes
almost independent of $n_B$ for $n_B\lsim 12n_0$, $m_\eta^*
/m_\eta \simeq 0.4$, within the $n_B -T$ region depicted in
Fig.~\ref{MetaT200}. For $n_B\gsim 15n_0$ the in-medium $\eta$
mass practically is independent of temperature. The $\eta$
condensation does not appear for $T\leq T_{c\sigma}$. If we used
parameter choice $\Sigma_{\eta N}=280$~MeV and $\kappa_{\eta
N}=0.4$~fm (see solid line in Fig. \ref{Meta}) we would meet with
the $\eta$ condensation problem.

\subsection{SHMC EoS for baryonless matter.}

Now let us consider the case $n_B \simeq 0$ that is close to
conditions realized at RHIC. The decrease of the hadron masses
with increase of the temperature for $n_B \simeq 0$ has been found
in~\cite{V04} as the consequence of  the blurring of the baryon
and  meson vacuum. Here we obtain a similar effect but within the
quasiparticle picture.

\begin{figure}[thb]
\hspace*{-2mm}\includegraphics[height=67mm,clip]{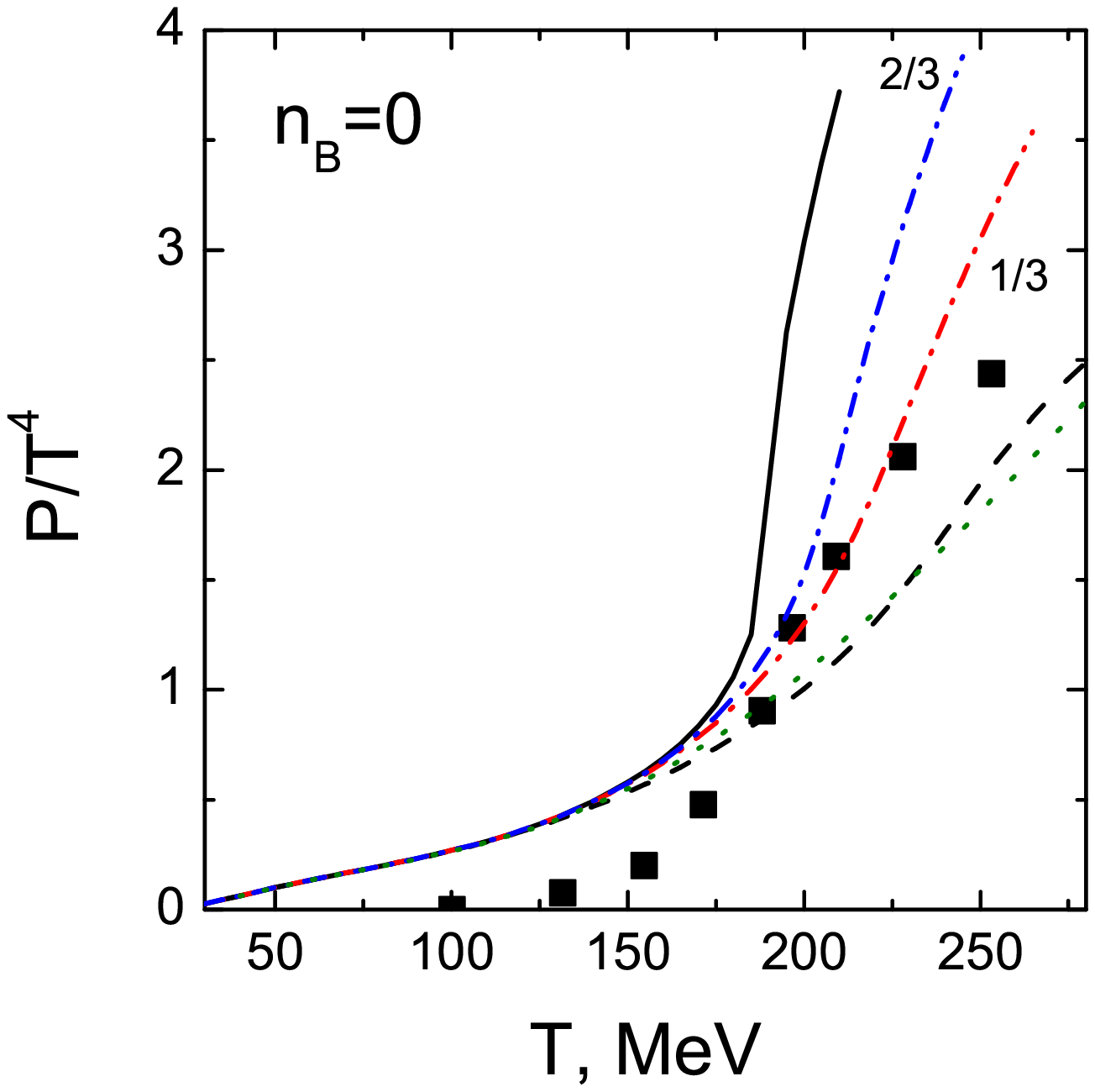}
\includegraphics[height=67mm,clip]{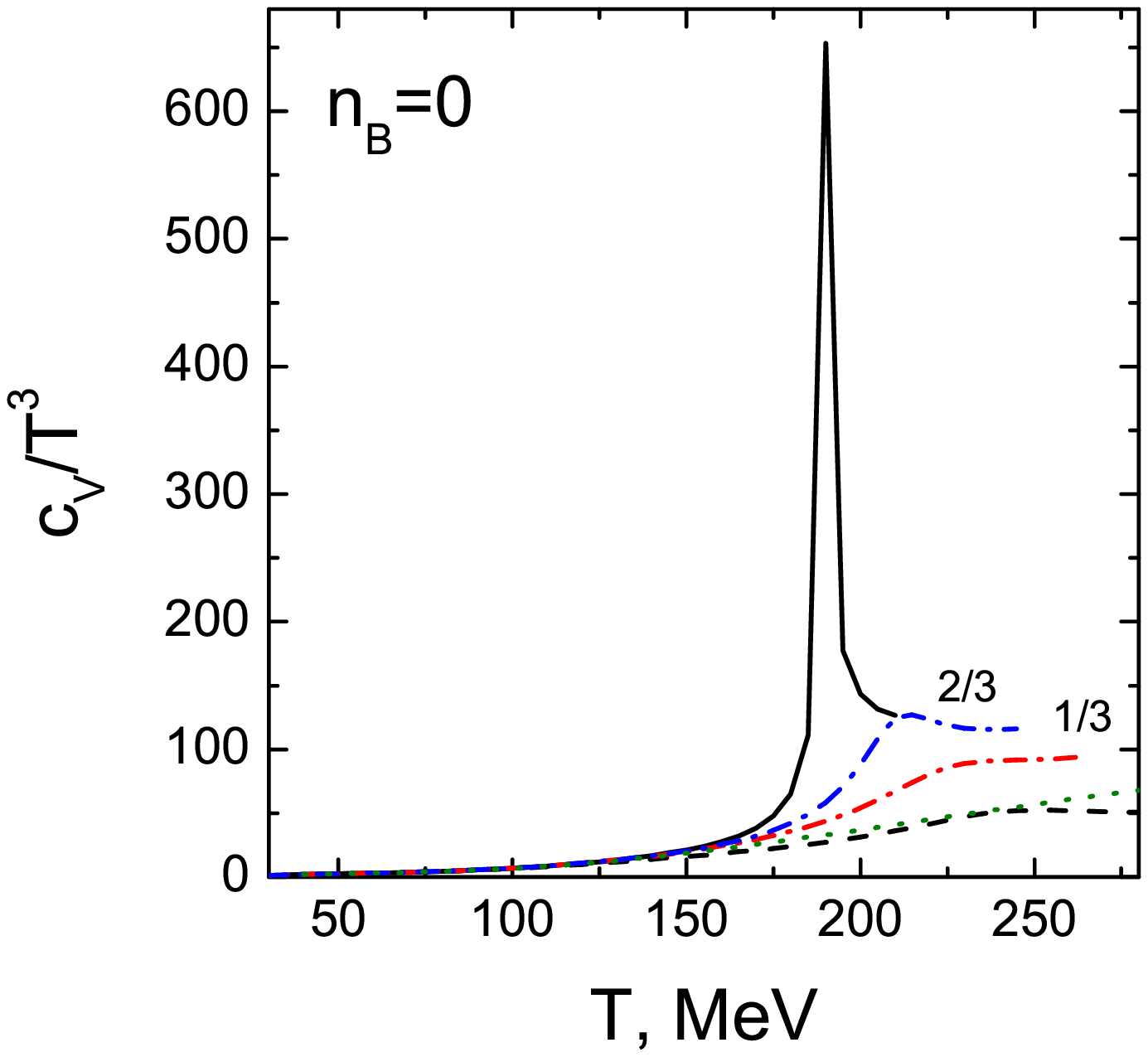}
\caption{ Temperature dependence of the reduced pressure (left)
and reduced specific heat (right) for iso-symmetric baryonless
matter. The solid line presents our full calculation. The dashed
line shows the case when contribution of all baryons besides
neutrons and protons is artificially suppressed. Two dashed-dotted
lines are the case when all $g_{mb}$ couplings except for nucleons
are suppressed by factors of $2/3$ and $1/3$, respectively.
Squares show  the lattice QCD result for the 2+1 flavor
case~\cite{KLP01}. The dotted curve is for the IG model.}
 \label{pT4}
\end{figure}
In Fig.~\ref{pT4}  temperature dependence of the pressure is shown
in $T^4$ units (left) and the specific heat, in $T^3$ units
(right) for $n_B =0$. On the left panel the solid curve presents
our calculation with parameters determined in sect. \ref{Det}. The
dashed curve shows pressure in the case when contribution of all
baryons, except neutrons and protons, is artificially suppressed
while the dotted curve is given for the IG model. All curves are
close to each other for $T\lsim 160\div 170$~MeV. Differences
start at higher temperatures where  high-lying baryon resonances
come into play. At $T=T_{c\sigma}$ the effective mass squared  for
the  $\sigma$ excitation drops to zero and becomes negative for
$T>T_{c\sigma}$. In order not to complicate the consideration we
avoid discussion of the hot Bose condensation of $\sigma$
excitations. Therefore we cut the solid line at $T>T_{c\sigma}$.
In the case when baryon resonance contributions (except nucleons)
are artificially suppressed, the effective nucleon mass as well as
the $\sigma$ excitation mass approach zero only at $T\rightarrow
\infty$ (if $\eta$ is a free particle, otherwise $\eta$ may
condense at $T_{c\eta} \simeq 265$~MeV).

We also  compare our result with the lattice QCD calculations for
$2+1$ flavor \cite{KLP01}. One usually believes the lattice data
for the quark-gluon sector (i.e. for $T>T_{\rm dec}\approx
175$~MeV and for $n_B \simeq 0$). On the other hand, there are
doubts that the lattice calculations produce appropriate results
for the hadronic sector, $T<T_{\rm dec}$, since they get
unrealistically high value for the pion mass, $m_\pi\sim 700$~MeV,
instead of the physical value $140$~MeV, cf. \cite{KLP01}.
Recently a higher deconfinement temperature was obtained in
lattice QCD calculations for similar system with almost physical
value of the pion mass, $T_{\rm dec}=192(7)(4)$~\cite{Um06}.
However thermodynamical characteristics still have not been
recalculated. Therefore we may use any hadron EoS for $T<T_{\rm
dec}$ (with a not precisely known value of $T_{\rm dec}$) without
referring to the lattice data. For a larger temperature, one could
expect that the quark phase becomes energetically preferable. It
happens if the pressure in the hadron phase is less than in the
quark phase. Oppositely, Fig.~\ref{pT4} shows that the pressure
extracted from the lattice calculations is significantly smaller
than that obtained within the SHMC model for the hadron phase at
$T\gsim 170$~MeV. This may mean that we have no deconfinement
phase transition with our EoS with the values of the coupling
constants used. Instead, within our model we obtain a state of  a
high-temperature hadron gas of many baryon resonances
(quasiparticles in the given model) and bosons, with small
effective  masses. This phase is enriched by antiparticles. As we
have mentioned above, this result coincides with that of
Ref.~\cite{V04}, though here it is obtained within a
phenomenological quasiparticle scheme while in~\cite{V04} it is a
consequence of blurring of the hadron vacuum.

The deconfinement transition could be constructed if
 the  (lattice) quark-gluon EoS is matched with our EoS for lower
$T$ ($T_{\rm dec}\lsim 160$~MeV)  when our hadron  phase pressure
is not yet too high (compare solid and dashed lines in
Fig.~\ref{pT4}). Another possibility for the deconfinment
transition can be associated with a different mechanism: The
overlapping of the hadron cores, if hadrons are considered as
composite particles. We have  a dramatic increase of hadron
degrees of freedom  at $T\gsim 170$~MeV. Thus the hadron cores may
become overlapping for such temperatures. If this mechanism works,
the deconfinment transition should be treated as an enforced
Mott-like transition occurring due to the melting of composite
hadrons, rather than matching the pressures of two different
phases.

In  Fig.~\ref{pT4}
solid lines present results of calculations with the default
parameters defined in sect. 4. Here we have $T_{c\sigma}\simeq
 210~$MeV. Dashed-dotted lines demonstrate the cases when $g_{mb}$ couplings,
except for nucleons, are suppressed by factors $2/3$ and $1/3$ (as
indicated on the plot). The latter  case (with $1/3$ prefactor)
allows to fit the lattice data up to $T\sim240~$MeV. In this case
a quark liquid would masquerade as a hadron one. In principle one
could fit the lattice data in a still larger region of
temperatures (e.g., up to $500~$MeV) introducing $\chi_\sigma
<\Phi$.  A violation of the universality of the $\sigma$ scaling
would be in a line with that we have used for $\om$ and $\rho$,
$\eta_{\om}\neq 1$ and $\eta_{\rho}\neq 1$. However we will not
elaborate this possibility in the present work. The dashed-dotted
curve labelled by $1/3$ is cut at $T_{c\eta}\simeq 265~$MeV. At
this point the $\eta$-meson mass becomes imaginary within our
parametrization and $\eta$ condensate arises. If $\eta$ were
treated as a free hadron, we would obtain $T_{c\sigma}\simeq
330~$MeV in this case.

The right panel of Fig.~\ref{pT4} demonstrates the temperature
behavior of the specific heat for $n_B=0$. With the standard
choice of couplings (solid curve) we observe a sharp peak with
maximum at $T\simeq 190~$MeV.  At this point the second derivative
of the effective nucleon-$\om$-$\rho$ and $\sigma$-excitation
masses changes the sign. The specific heat retains a continues
function. The  obtained behavior is typical for the strong
crossover transition. Note that such a behavior of the specific
heat was also found within the standard RMF model~\cite{Theis83}.
In the case of the second-order phase transition, the specific
heat would be discontinues at the critical point. When $g_{mb}$
couplings are suppressed, the peak is smoothed that reminds about
a weak crossover. Often the temperature at the maximum of the
specific heat is associated with the critical temperature of a
phase transition. However note that in our case the position of
this maximum  $T\simeq 190~$MeV slightly differs from the
$T=T^{\rm chir}$ and it differs also from $T_{c\sigma}$ ($\sim
210~$MeV), see Fig.~\ref{Mz}. This fact of non-unique value of the
critical temperature $T_c$ is also manifested in the lattice
calculations: Analysis of different thermodynamic quantities leads
to different numerical values of $T_c$ even in the continuum and
thermodynamic limits~\cite{AFKS06}. One should keep in mind that
there is no liberation of internal (quark-gluon) degrees of
freedom of hadrons in the RMF models. Note that the $P/T^4$ curve
calculated in the IG model is below our result even if only
nucleons are taken into account (compare dotted and dashed curves
in left panel of Fig.~\ref{pT4}). This is due to the decrease of
the nucleon mass with  increasing temperature in the SHMC model.
The specific heat $c_V /T^3$ in the IG model does not saturate for
high $T$, whereas all curves of the SHMC model tend to constant
values.

\subsection{SHMC EoS for $n_B \neq 0$}

\begin{figure}[thb]
\hspace*{-2mm}\includegraphics[width=70mm,clip]{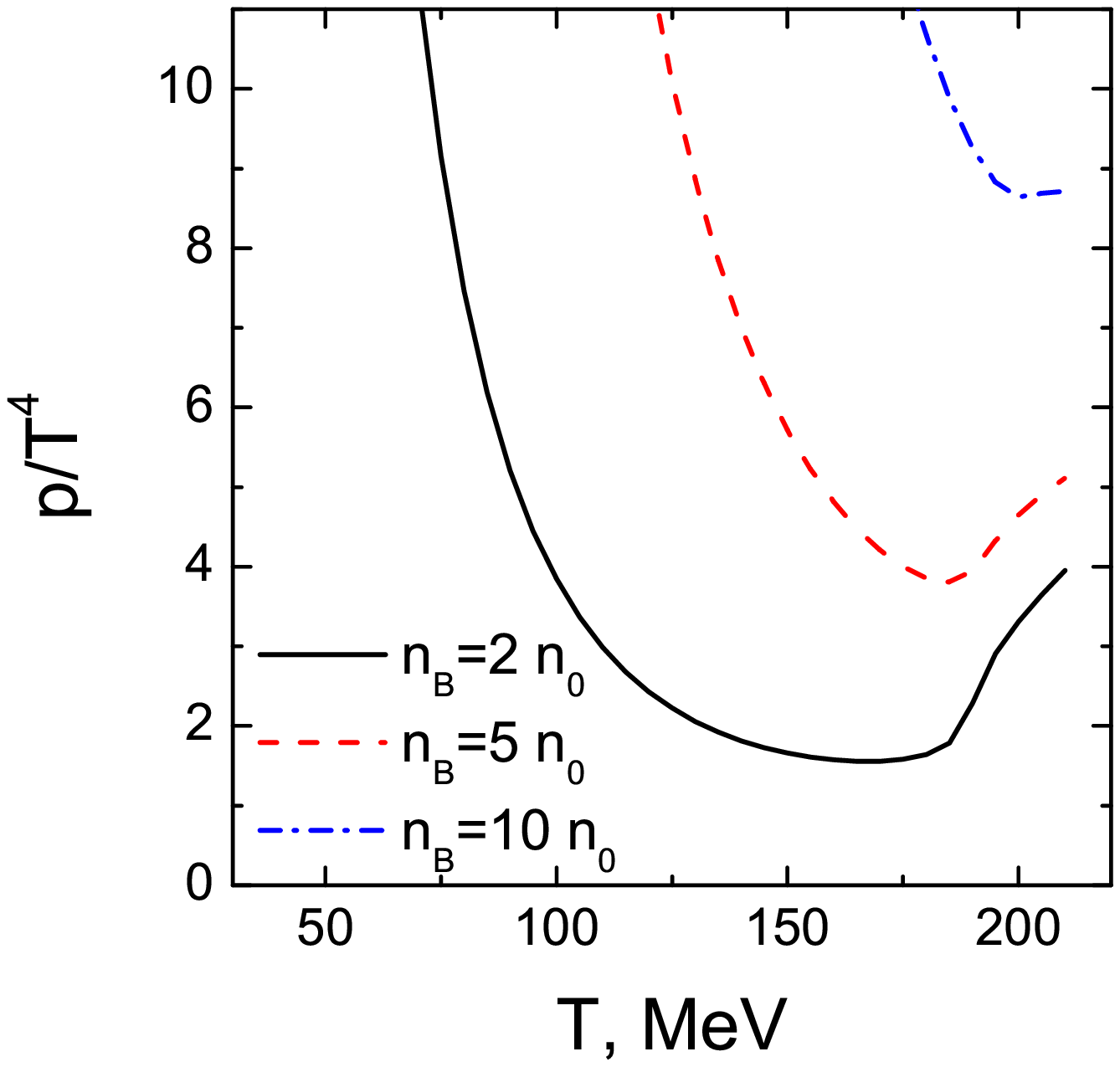}
\includegraphics[width=70mm,clip]{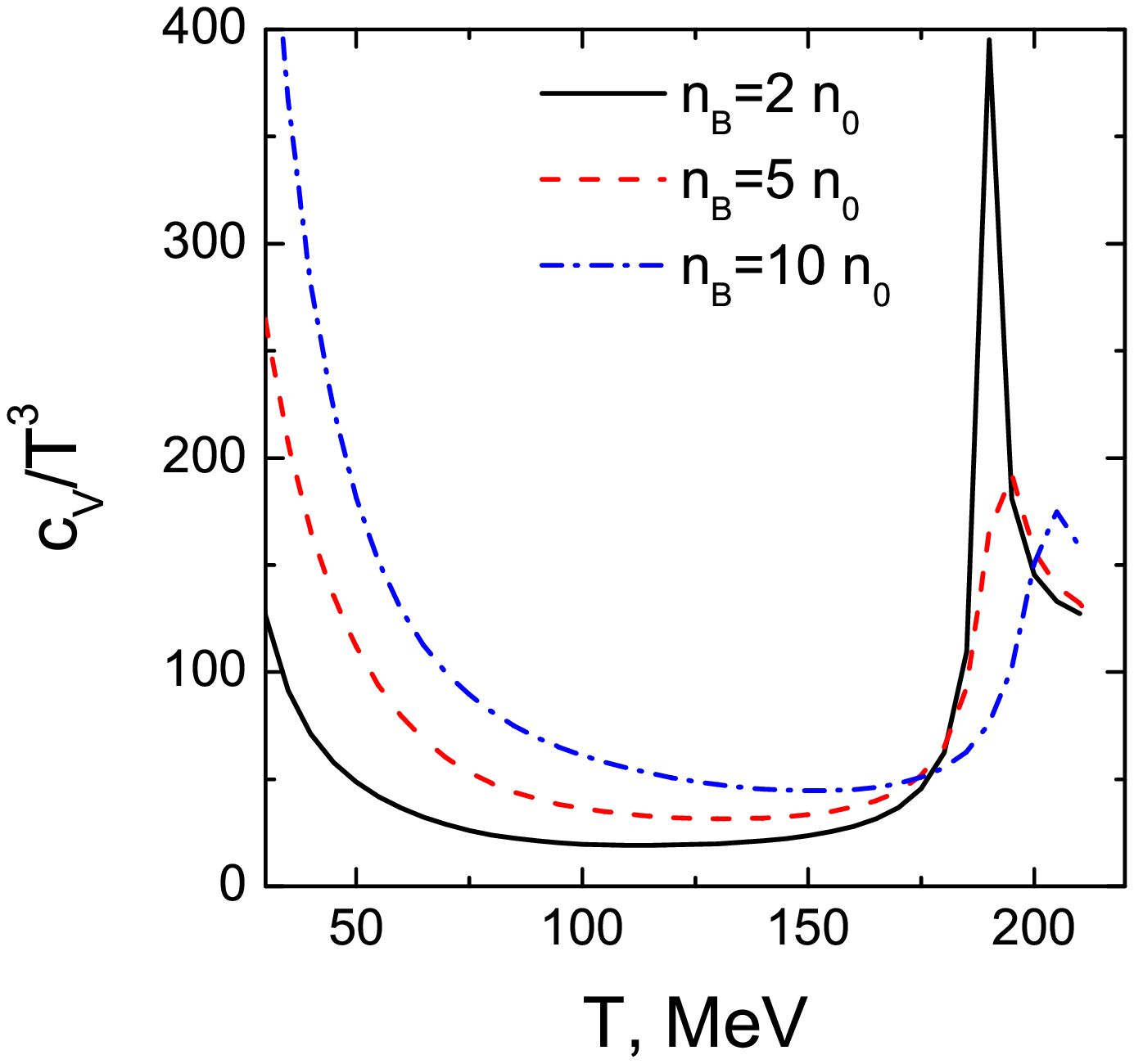}\caption{ Temperature dependence
of the pressure (left) and specific heat
  (right) at  $n_B \neq 0$ for $N=Z$ for three values of $n_B$.
 }
 \label{pT4r}
\end{figure}

In Fig.~\ref{pT4r} we show temperature dependence of the pressure
(left) and specific heat (right) for  baryon densities $n_B/n_0
=$2, 5 and 10.
 The pressure gets minimum at temperature $T\simeq
(175\div 200)~$MeV and grows for higher $T$. The calculations are
stopped at $T=T_{c\sigma}$ as was mentioned above. The value
$T_{c\sigma}$ depends on $n_B$ only moderately. We see that the
peak of the specific heat survives at finite baryon density. The
value of the temperature $T_{\rm max}$ corresponding to the peak
is slightly shifted up with the baryon density  but the hight of
the peak changes significantly. In particular, the position of the
$c_v/T^3$ maximum is $T_{\rm max}\sim 190~$MeV at $n_B\lsim 2n_0$
and moves to about 200~MeV for the density $n_B= 10n_0$.

\subsection{Particle densities}

The particle density of species $i$ is given by
\be \label{dens}
n_i=g_i \int \frac{d^3p}{(2\pi)^3} \ f_i(p)~,
\ee
where the
spin-degeneracy factor $g_i=(2s_i+1)$ and $f_i$ is the particle
occupation function, see
Eqs.~(\ref{oc})-(\ref{Tmu}),(\ref{Tbos}),(\ref{Tmubos}).
 In the SHMC  model
the spectra of most particle species are getting softer with the
baryon density increase because of in-medium effect. This occurs
for all baryons and for $\om$ and $\rho$ due to the scaling of the
effective masses, as well as for $\sigma$ and (quasi)Goldstone
excitations ($K^-$, $\bar{K}^0$ and $\eta$) as a consequence of
their interaction with mean fields. Therefore the densities of
these particle species are larger than in the IG case.

\begin{figure}[thb]\hspace*{-10mm}
\includegraphics[width=150mm,clip]{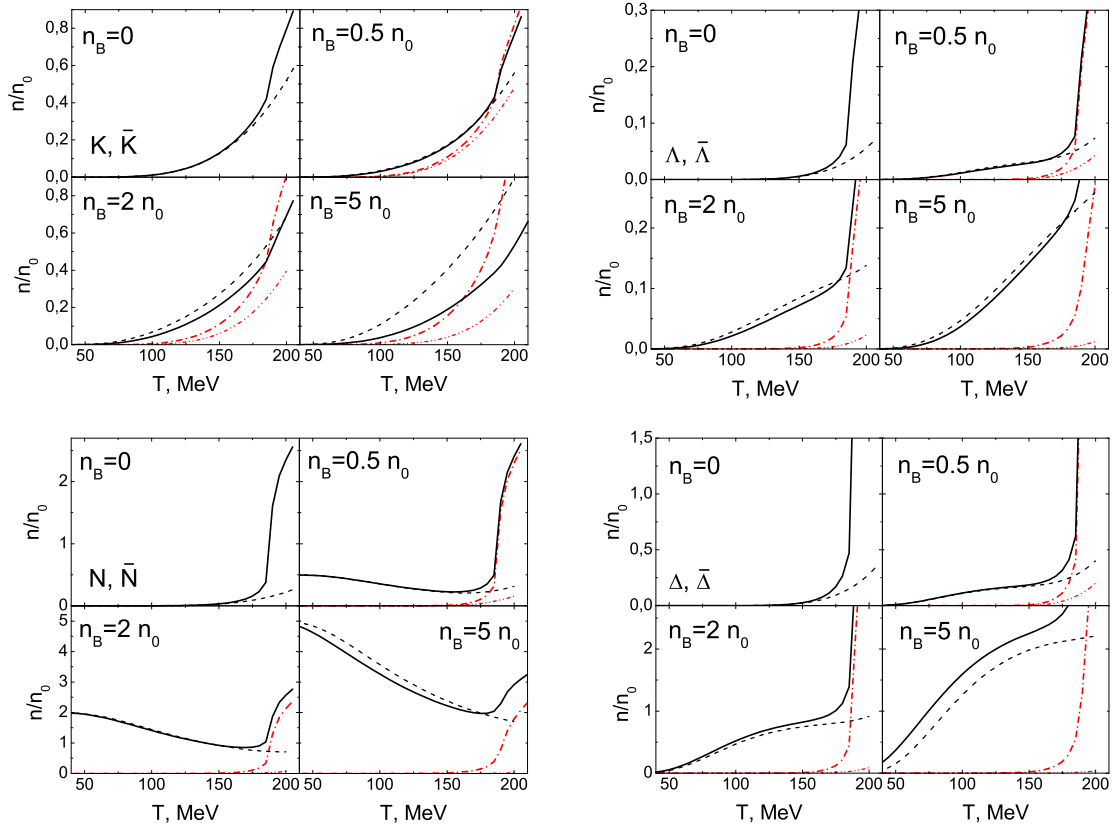}
\caption{ Particle number density in $n_0$ units versus
temperature  at different baryon densities for $K, \Lambda, N,
\Delta$ and their antiparticles, $N=Z$. Our model  results are
plotted by solid line for particles and by dashed-dotted line for
antiparticles. Dashed and  dashed-double-dotted curves correspond
to particles and antiparticles  in the IG model. }
 \label{npart}
\end{figure}
 Temperature dependence of the density for
various particle species  (and for their antiparticles)  for
iso-symmetric matter at different values of baryon densities is
shown in  Fig.~\ref{npart}.  In the $n_B =0$ case the particle and
antiparticle yields coincide. As it is seen, all the species
besides nucleons exhibit very similar behavior (compare solid and
dashed lines). Particle number density slowly grows with the
temperature increase till $T\sim 170~$MeV and then rapidly goes
up. In the range $T\lsim 170~$MeV the SHMC results  are rather
close to those of the IG model  (except for a high density, see
example $n_B =5n_0$) but drastically diverge at higher
temperatures. This difference is naturally explained by the rapid
decrease of in-medium masses in the SHMC model at these
temperatures (see Figs.~\ref{Mz0},\ref{Mz}). In contrast, at
$T\lsim 170~$ MeV the nucleon density $n_N$ stays almost
independent of temperature at $n_B\simeq 0.5n_0$, or significantly
decreases at higher $n_B$. These facts are due to a stabilization
effect of the baryon conservation law and a strong growth of
production of baryon excited states and hyperons with the baryon
density  increase (cf. $\Delta$ and $\Lambda$ particle densities
in Fig. \ref{npart}). Similarly to other species, the nucleon
particle density rapidly goes up with further temperature increase
(for $T\gsim 170~$MeV).

Temperature-density behavior for antiparticles is quite different.
In spite of the fact that  the iso-symmetric system is considered,
even for IG the densities of created kaons and antikaons are
different (except for the $n_B =0$ case). This is a consequence of
the total strangeness conservation which takes into account also
hyperon species. Thus $\mu_{\rm str}$ proves to be non-zero even
for $N=Z$ that results in $n_{K^+}\neq n_{K^{-}}$. Although the
correction to the energy dispersion curve is positive (repulsion)
for kaons and negative (attraction, see Fig.~\ref{omKocmp}) for
antikaons, their number densities intersect at temperature
$160\lsim T\lsim 180~$MeV for $n_B\neq 0$ and then, with the
subsequent temperature increase, the antikaon density even exceeds
that for kaons. But it is only apparent violation of strangeness
conservation since excited kaon states $K^*$ contribute. Because
$K^*$ is treated as a free particle, $n_{K^{*+}}\gg n_{K^{*-}}$ at
high temperature and baryon density. In particularly, at
$T=190~$MeV and $n_B=5n_0$ we have $n_{K}=0.46n_0$, $n_{\bar
K}=0.68n_0$ and for the excited kaon state $n_{K^*}=1.15n_0$,
$n_{\bar{K}^*}=0.07n_0$. Indeed, $n_{K}<n_{\bar K}$ but the total
strangeness of these four kaon species is $\sim 0.8n_0$ which is
compensated by hyperons. As for antinucleons, antideltas and
antihyperons, their behavior looks very similar: the yield of all
antibaryons is markedly suppressed at $T\lsim 170~$MeV and
abruptly grows up at higher temperature, $T\gsim 180~$MeV. This
increase is not reproduced by the IG model which produces
significantly  lower hadron densities. Generally, the baryon
density dependence of the particular number density seems to be
not as strong as temperature one.

\section{Application of the model to HIC}\label{hic}

\begin{figure}[thb]
\hspace*{3mm}
\includegraphics[width=120mm,clip]{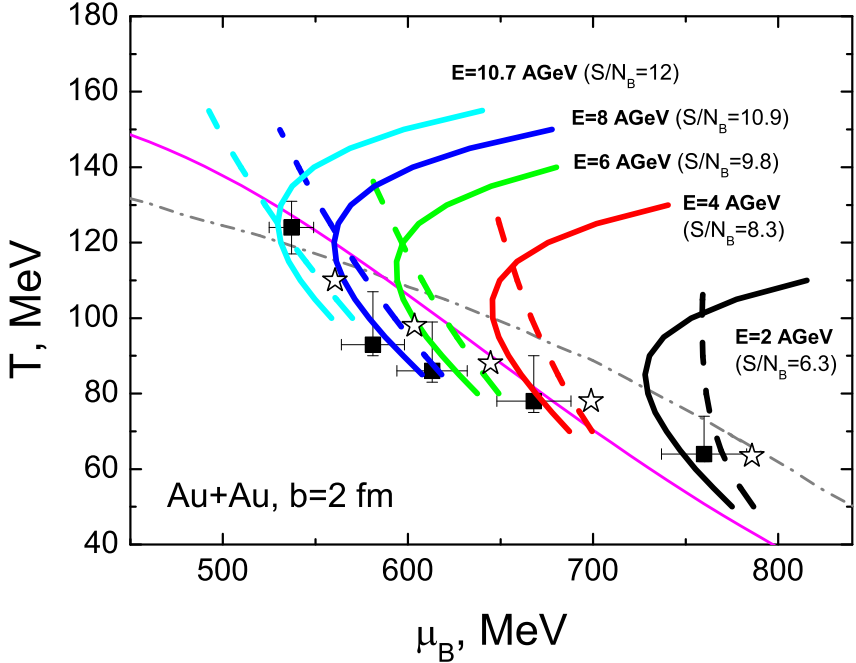}
\caption{ Isentropic trajectories for central Au+Au collisions at
different bombarding energies calculated in our model (solid
lines) and IG (dashed lines). Experimental points with error bars
are taken from~\cite{ABMS05}. The freeze-out points marked by
stars are obtained in this work (see the text below). Thin line
corresponds to the freeze-out curve in~\cite{ABMS05} while
dash-dotted line is that from~\cite{CR98}. }
 \label{isoent}
\end{figure}

The SHMC model describes  the EoS of hot and dense hadronic matter
in a  broad  range of temperatures and baryon densities. In HIC
the dense matter created in the initial stage  is expected to
rapidly thermalize and then it expands without significant
generation of the entropy $S$. In the process of expansion some
particles may leave the fireball carrying away a part of the
entropy. Thus more appropriate characteristic which is
approximately conserved is the entropy per participant nucleon
($S/N_B$). This thermodynamic quantity should be conserved in
quasi-equilibrium case and it is also less affected by any
possible particle loss or gain from the fireball during the
expansion stage. The predictions of our models for the evolution
path in the $(T,\mu_B)$-plane as obtained from the SHMC EoS under
condition of the fixed $S/N_B$ are shown in Fig.~\ref{isoent}.

For the initial fireball expansion stage,
 in the case of semi-central Au+Au collisions at different
bombarding energies below the top AGS energy, the reduced entropy
ratios,  $S/N_B$, were estimated within a transport Quark-Gluon
String Model (QGSM), as described
in~\cite{TS}\footnote{{As shown recently
\cite{ABC06} dynamical trajectories described by the QGSM are
quite close to those in the UrQMD model providing the closeness of
initial fireball stages in these two approaches.}}. The temporal
$S/N_B$ dependence exhibits some saturation \cite{TCN01}, values
being taken as an input for our isentropic calculations and
indicated in Fig.~\ref{isoent}. {The presented
isentropic curves follow exactly and unambiguously from the EoS.
Some uncertainties in location of these curves are coming from
uncertainties in the estimate of the reduced entropy related to
the bombarding energy. In our case $\delta (S/N_B)\sim 0.1$ and it
unessentially influences the trajectory location. } The
$(T,\mu_B)$ trajectories calculated within the SHMC model show
turning points those positions correlate roughly with the
freeze-out curve. This fact was noticed earlier in~\cite{TCN01}
where the EoS with a phase transition was used. There the
high-temperature part of trajectory with $\partial T/\partial
\mu_B >0$ was associated with the quark-gluon sector of EoS but in
the present work it is due to a strong decrease of hadron masses
in the considered area of the phase diagram. Note that there is no
turning point in the IG case.

Two chemical freeze-out curves are shown in Fig.~\ref{isoent}.
Dashed-dotted curve corresponds to the IG EoS with the chemical
freeze-out condition that the energy per hadron equals to 1 GeV
($<E>/<N>=1$ GeV)~\cite{CR98}. {It is of interest
that in the considered $(T,\mu_B)$ range this freeze-out curve is
also very close to that obtained for the IG EoS with the fixed
baryon density $n_B=0.12 \,$ fm$^{-3}$~\cite{ABMS05}. So, the net
baryon density of states above the dashed-dotted line in
Fig.~\ref{isoent} is $n_B>0.8n_0$. Maximal densities explored in
this phase diagram range from $\sim 2.3n_0$ to about $4n_0$ when
the bombarding energy increases from 2 AGeV till the top AGS
energy (these numbers depend on the region over which averaging
was made). } The thin line is obtained by interpolation of the
$(T,\mu_B)$ fitting parameters extracted at every available
bombarding energy by the $\chi^2$ minimization of the difference
between experimental and theoretical (calculated within a
statistical model with the IG EoS) hadron abundance~\cite{ABMS05}.
In any analysis the thermodynamic quantities $(T,\mu_B)$ at the
freeze-out are derived from the analysis of measured particle
ratios. The straightforward consequence of the statistical
assumption is that the mean hadron multiplicities should be
calculated in the full ($4\pi$) phase space. However, due to
limited experimental acceptance it is not possible to do that in
each case and instead the particle ratios at the middle rapidity
are used. Making use of  middle rapidity ratios implies  that the
measured distribution $dN/dy$ is approximately constant over the
same range. The analysis of the expected dispersion of rapidity
distributions shows that the use of full space multiplicities  is
better suited over the energy ranges of AGS and SPS~\cite{BMG06}.
Unfortunately in the energy range considered in Fig.~\ref{isoent}
the $4\pi$ particle ratio measurements are available only at two
energies, and  the thin line is obtained by the $(T, \mu_B )$
interpolation based on the analysis of the middle rapidity data.
Generally, both heuristic freeze-out curves describe quite well
the extracted
 $(T,\mu_B)$ values   and they differ more noticeably in the
presented range $E_{lab}\lsim 10~$AGeV, below the top of AGS
energy.  Before to explain how freeze-out points marked by stars
in Fig.~\ref{isoent} were obtained in the given work, we consider
which hadron abundance is predicted by the SHMC model and which
values of thermodynamic parameters may be inferred  from
comparison with experiment.

\begin{figure}[thb]
\hspace*{3mm}
\includegraphics[width=130mm,clip]{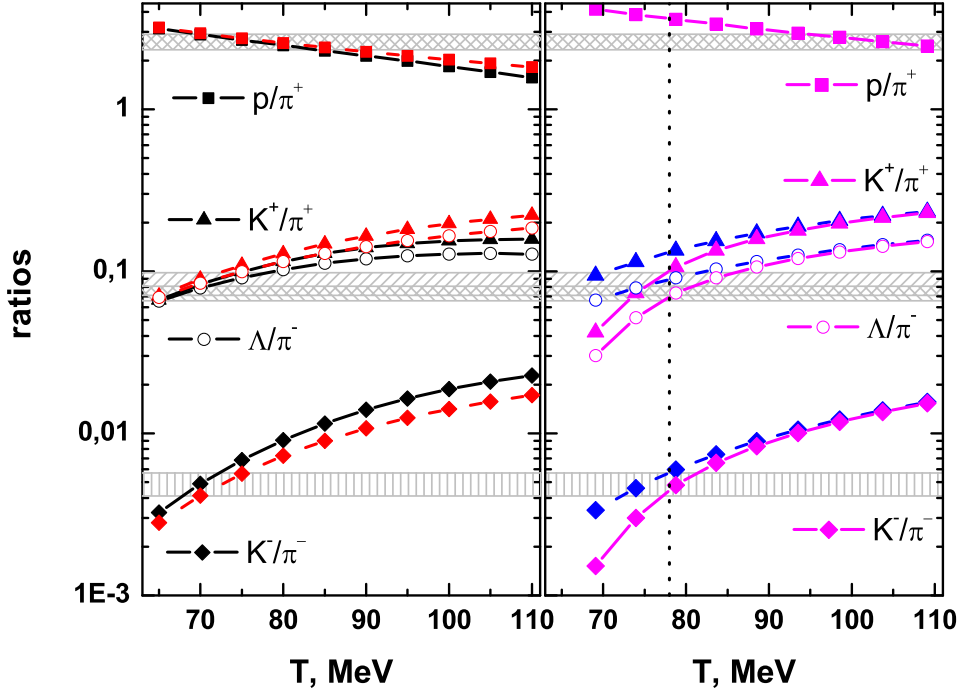}
\caption{ Particle ratios calculated along the isentropic
trajectories for central  Au+Au collisions at $E_{lab}=4$ AGeV.
Left panel: Solid and dashed lines are calculated right before
freeze-out for the SHMC and IG EoS, respectively. Right panel:
Solid line is calculated in the SHMC  model right after freeze-out
has occurred, with taking into account the canonical strangeness
suppression. Dashed line is that without strangeness suppression.
Shaded bands correspond to uncertainties of experimental
data~\cite{ABMS05}. The dotted vertical line is our estimate of
the freeze-out temperature.}
 \label{ratio4}
\end{figure}
In Fig.~\ref{ratio4} the $E_{lab}=4$ AGeV case is exemplified in
detail. In the left panel the particle ratios for the most
abundant hadrons are calculated along isentropic trajectories of
the SHMC and IG EoS's depicted in Fig.~\ref{isoent}. Note that
these quantities are {\em{in-medium ratios}} which would be
"inside" hot and dense matter right before the freeze-out, but
measured ratios correspond to free particles after the freeze-out.
Although trajectories are remarkably different for the IG and SHMC
EoS's, the particle ratios turn out to be not so sensitive to the
choice of EoS (besides the $K^-/\pi^-$ ratio), since freeze-out
baryon densities are rather low. In the right panel of
Fig.~\ref{ratio4} the particle ratios are shown just after the
freeze-out.

Up to now there is no appropriate theory of the freeze-out though
there exist many different recipes. Generally, transition from the
collective expansion to kinetic stage and then to free particle
streaming should be continuous and it takes some finite time. For
the sake of simplicity some sudden approximation is often applied.
If we assume a prompt freeze-out concept, the observable yield
will be defined by in-medium particle spectra $dN/d^3 p$, as we
calculated them within the SHMC model,  but additionally
multiplied by a pre-factor $\sqrt{m^2 +p^2}/\sqrt{m^{*2} +p^2}$
due to quasiparticle undressing, cf. \cite{VS87}. However the
assumption of prompt freeze-out might not be applicable at least
for some particle species. Besides, one should take a especial
care about the total energy conservation.

\begin{figure}[t]
\hspace*{1mm}
\includegraphics[width=140mm,clip]{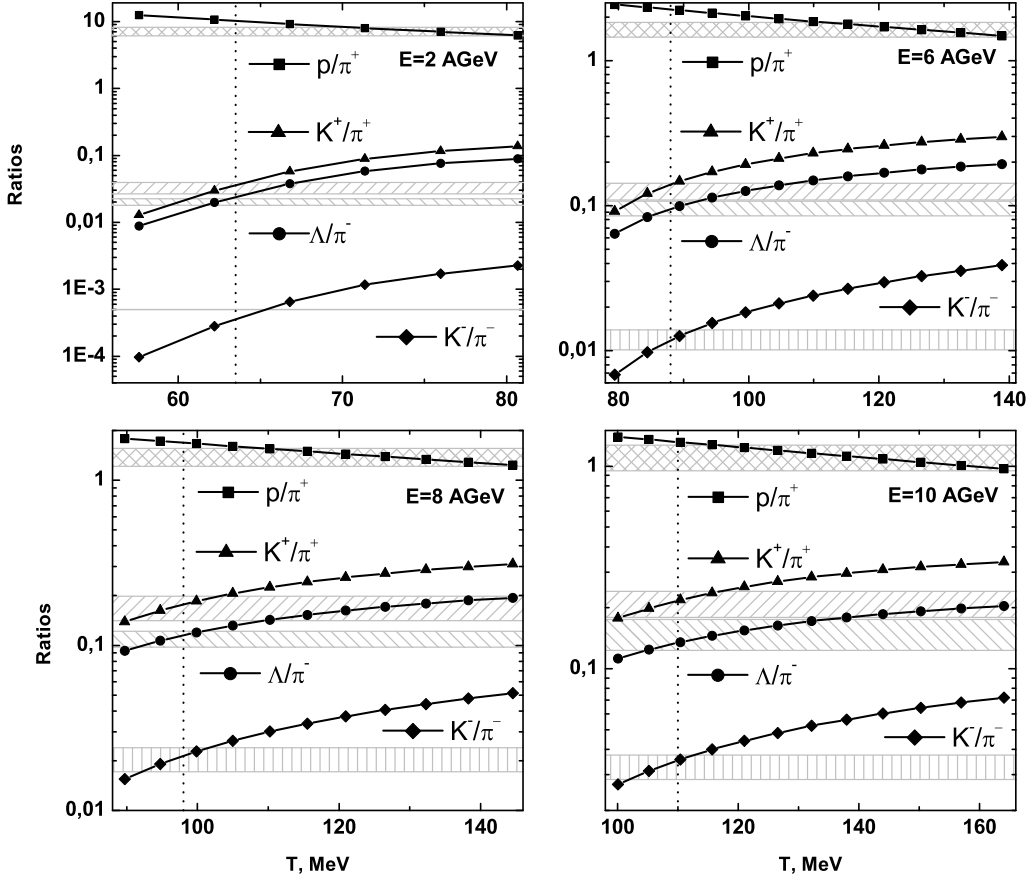}
\caption{ Particle ratios for central  Au+Au collisions calculated
at different collision energies. Solid lines are SMHC results at
the shock-like freeze-out with taking into account canonical
strangeness suppression. Shaded bands correspond to uncertainties
of experimental data~\cite{ABMS05}. In the case of
$E_{lab}=2~$AGeV the experimental line for $K^-/\pi^-$ shows only
the upper limit of this ratio. The dotted vertical lines are
freeze-out temperatures.}
 \label{allH}
\end{figure}

In Ref.~\cite{Bug} another more conventional choice was suggested.
At crossing the freeze-out point (hypersurface in a general case)
the change from the in-medium to IG  EoS occurs in a "shock-like"
way: One  demands the energy, momentum, and the net baryon charge
and strangeness conservations. We do not consider the fireball
expansion dynamics but use the isentropic trajectory  with our
EoS. Thus we know nothing about momentum conservation and  in our
simplified case only conservation of energy and charges is taken
into account. Certainly, important collective flow effect is out
of consideration. Therefore below we focus on analysis of particle
ratios, where this effect  is cancelled, cf.~\cite{CR98,ABMS05}.
We use this scenario~\cite{Bug} in our calculations presented in
Figs.~\ref{ratio4} (right panel) and Fig.~\ref{allH}. In this case
the attractive in-medium interaction brings to an increase of
temperature of free gas after freeze-out.  As shown in the right
panel of Fig.~\ref{ratio4} (dashed lines), such a procedure
results in the agreement with experiment of particle ratios  at
temperature by about 10 MeV higher than that inside the medium
(cp. the left panel of Fig.~\ref{ratio4}). Certainly, for the IG
EoS there is no changes due to the energy-momentum conservation.

All above-mentioned calculations have been done in the grand
canonical ensemble. However for description of strangeness
production at not too high temperatures when a number of strange
particles is small, this approach is not quite appropriate and the
canonical ensemble for strangeness should be used~\cite{Redlich}.
Replacement of the grand canonical description by the strangeness
canonical one results in extra  temperature-dependent
suppression factor for strange particle densities which reaches unity
for $T\gsim 100~$MeV. The factor was calculated in the standard
way~\cite{ABMS05,TP05}. This canonical strangeness suppression
effect is clearly seen from comparison of solid and dashed lines
in Fig.~\ref{ratio4} (right panel). Note that, if both effects are
taken into account, the best agreement of calculated particle
ratios simultaneously with all measured ones is reached at the
temperature $T\simeq 78~$MeV which can be treated as a freeze-out
temperature $T_{fr}$ in the given case (shown by the vertical
line).

Final SHMC results for hadron ratios at other four AGS energies
are presented in Fig.~\ref{allH}. Calculated along isentropic
trajectories, these  ratios take into account both the shock-like
freeze-out and canonical suppression effect (compare with
in-medium ratios shown in Fig.~\ref{ratio4} for the case
$E_{lab}=4$ AGeV). One can see that at every bombarding energy it
is possible to fix some freeze-out temperature $T_{fr}$ by
condition of the best agreement of the calculated ratios with
experiment. The extracted $T_{fr}$ are shown in
Figs.~\ref{ratio4}, \ref{allH} by vertical dashed lines. One
should note that the proton yield includes both direct protons and
those feeding from the resonance decays. However in the
intermediate energy range considered, a part of protons is bound
into light complex particles ($d, \alpha$). This nucleon
coalescence effect is the higher, the lower the bombarding energy
 is, and this effect is disregarded in our model. Thus a
discrepancy with experiment for $p/\pi^+$-ratios seen in upper
panels of Fig.~\ref{allH} should not be taken too seriously.

If the $T_{fr}$ on the isentropic curve is known, an appropriate
baryon chemical potential for the IG EoS at the freeze-out,
$\mu_{fr}$, can be found. These pairs of $(T_{fr},\mu_{fr})$ are
indicated by stars in Fig.~\ref{isoent}. The new freeze-out
points, based on the same middle rapidity particle ratios,
slightly differ by some shift in the chemical potential from those
obtained by direct fitting of these ratios in statistical
theory~\cite{ABMS05}. As to $T_{fr}$, our freeze-out temperature
for $E_{lab}=10$ AGeV is noticeably below the fitted one. In
contrast with lower energies, at the top AGS energy the set of 10
particle ratios was measured and used in the statistical model
fit, while our results are based on 4 ratios, as shown in
Fig.~\ref{allH}. If the $d/p, \bar{p}/p, \bar{\Lambda}/\Lambda,
\varphi/K^+$ ratios are excluded from this analysis, then one gets
$T_{fr}=108\pm 9~$MeV and $\mu_B=555\pm 18~$MeV~\cite{ABMS05} what
is in a reasonable agreement with our result (see stars in
Fig.~\ref{isoent}). Note that this new method for deriving
freeze-out points in the phase diagram allows one to keep some
memory on the collision dynamics (the value of the entropy per
baryon and isentropic trajectories inherent to the final expansion
stage) and takes into consideration the in-medium particle
modification. The freeze-out point is taken on the phase
trajectory which depends on the EoS as is seen from comparison
between the SHMC and IG models in Fig.~\ref{isoent}. Note that
partial ratios are sensitive to the choice of the freeze-out
scenario. If we used the prompt freeze-out concept~\cite{VS87} we
would obtain other yields.

\begin{figure}[thb]
\hspace*{3mm}
\includegraphics[width=120mm,clip]{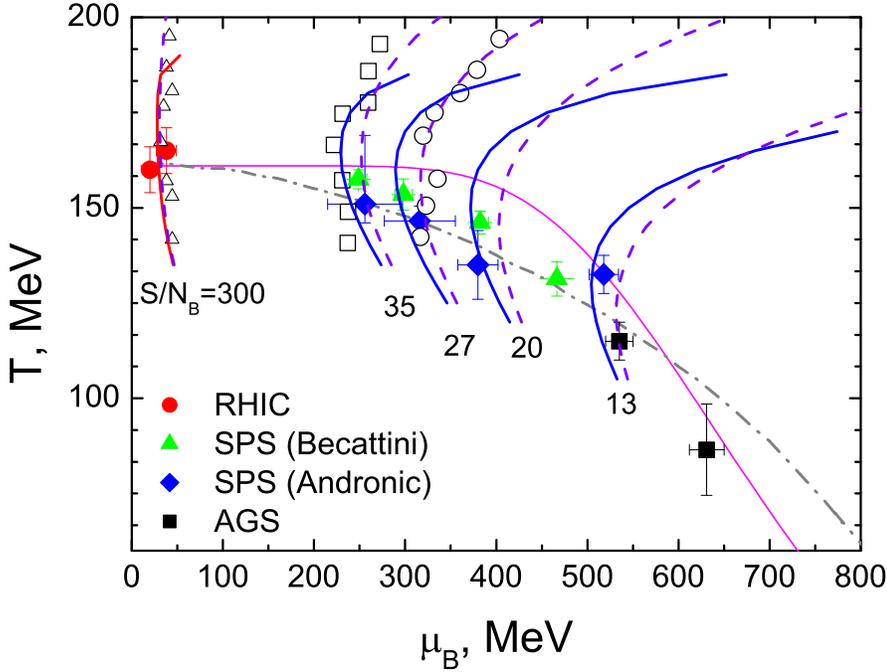}
\caption{ Isentropic trajectories for ultrarelativistic central
Au+Au collisions at different bombarding energies calculated in
the SHMC model with the default set of parameters (solid lines)
and with the suppressed  couplings by the factor 1/3 for all
baryons besides nucleons (dashed lines). Filled diamonds and
triangles are obtained from the $4\pi$ particle ratios
in~\cite{ABMS05} and \cite{BMG06}, respectively. Filled circles
are RHIC data based on the middle rapidity particle
ratios~\cite{ABMS05}. Open circles, squares and triangles are the
lattice 2-flavor QCD results~\cite{EKLS05} for $S/N_B=$ 30, 45 and
300, respectively. Two freeze-out curves are the same  as in
Fig.~\ref{isoent}. }
 \label{traekt}
\end{figure}

Let us extend our analysis to higher bombarding energies.  The
entropy per baryon participants was calculated in~\cite{RDB98}
within the 3-fluid hydrodynamic model assuming occurrence of
 the first order phase transition to a quark-gluon plasma. The energy
range from AGS to SPS was covered there. We use the $S/N_B$ values
for $E_{lab}=$158, 80, 40 and 20 AGeV, at which the particle
ratios were measured by the NA49 collaboration
(cf.~\cite{ABMS05}). At the RHIC energy we put $S/N_B=300$ in
accordance with the estimate in~\cite{EKLS05}. Note that in this
analysis the
 $(T,\mu_B)$ thermal parameters were determined using the
$4\pi$ particle ratios, besides the RHIC point. The difference
between two sets~\cite{ABMS05,BMG06} of $(T,\mu_B)$ freeze-out
points is caused by more elaborated statistical model implemented
in Ref.~\cite{BMG06}. It is of interest that both sets strongly
correlate with the freeze-out curve $<E>/<N>=1$ GeV~\cite{CR98}
and markedly differ from interpolation of the freeze-out points
based on the middle rapidity particle ratios.

The calculated trajectories for isentropic expansion are presented
in Fig.~\ref{traekt}. The trajectories (solid lines) behave very
similar to those at lower bombarding energies (shown in Fig.
\ref{isoent}) exhibiting some  flattening above the freeze-out
curve near the expected phase boundary. Such a behavior reminds
the one obtained in \cite{TNFNR03} where  a smeared first-order
phase transition is assumed. If baryon couplings (besides
nucleons) are suppressed by a factor 1/3, the calculated
trajectories turn out to be very close to the lattice QCD
results~\cite{EKLS05}. Note that dealing with the finite chemical
potential we use here the same choice of parameters $g_{mb}$ as in
the case $n_B=0$ (see Fig.~\ref{pT4}).

In principle, we could repeat our particle-ratio analysis  in the
SPS energy domain for both parameterizations of the SHMC model,
with suppressed and not suppressed couplings $g_{mb}$. However for
this aim the used SHMC model basis of hadron species should
noticeably be enlarged by inclusion of higher resonances since
available set of experimental data for particle ratios is
significantly larger at high energies~\cite{ABMS05}.

\section{
Conclusions and perspectives}\label{Concl}

In this paper the modified relativistic mean-field
$\sigma$-$\om$-$\rho$ model with scaled hadron masses and
couplings (SHMC model) formulated  for $T=0$  in Ref.~\cite{KV04}
is generalized to finite temperatures. Besides nucleon and mean
fields the model includes low-lying baryon resonances and their
antiparticles, boson excitations (following the $SU(3)$ concept)
and $\sigma$- $\omega$- $\rho$-excitations on the ground of mean
fields. The EoS for $T=0$ satisfies general constraints known from
atomic nuclei, neutron stars  and those coming from the flow
analysis of HIC data. Like in the KVOR model~\cite{Army}, we
assume that $\sigma$, $\om$, $\rho$ field mass terms, as well as
nucleon masses, decrease with increase of a combination of a
$\sigma$ mean field $f=g_{\sigma N}\chi (\sigma ) \sigma /m_N$
(see Eq.~(\ref{f})) corresponding to the change of the chiral
condensate density.  The model supposes the simplest choice of the
Brown-Rho scaling when the change of all masses mentioned above
follows  the same universal law. In order to describe properly the
EoS,  a similar scaling of the coupling constants is introduced
(with a slight violation of the universality of the scaling law).


 It was shown that at $T=0$ our model simulates the
(partial) chiral symmetry restoration with the baryon density
increase. The baryon-$\om$-$\rho$ and $\sigma$ excitation masses
fall down with the baryon density increase in the interval $0<n_B
<n_{min,B}\simeq 8n_0$. To certain extent, this is naturally
predetermined by the imposed Brown-Rho scaling, in terms of
the factor $\Phi=1-f$ (see
Eqs.~(\ref{bar-m}) - (\ref{Br-sc})). For higher densities the
masses begin to grow up. Such a behavior resembles the
density-dependence of chiral and gluon condensates obtained in
Refs.~\cite{BW,RK}.

Although the Lagrangian of the SHMC model does not respect chiral
symmetry the model  simulates a chiral symmetry restoration with a
temperature increase. In a narrow interval of temperatures
$170\lsim T\lsim 210$~MeV the nucleon-$\om$-$\rho$ and $\sigma$
excitation masses drop to zero (at $T_{c\sigma }\simeq 210$~MeV)
for the case $n_B =0$ and similarly for finite $n_B$. Masses of
higher lying resonances also fall down but do not vanish at
$T_{c\sigma }$.

Thus at high temperatures and densities we deal with strongly
interacting hadron matter which exhibits some (pre)critical
properties. These properties can be related to the decrease  of
the $\sigma$ effective mass. The $\sigma$ excitation mass attains
zero with the temperature increase till $T_{c\sigma}$, while it
never decreases below a half of the bare mass at $T=0$. The
temperature dependence of the EoS at $T$ near $T_{c\sigma}$
reminds the "phase boundary" behavior, far away from the IG EoS.
Particularities of the thermodynamic behavior near this "phase
boundary" at high temperature were discussed in detail. In the
case $n_B =0$ the SHMC model results are compared with the lattice
QCD data. For $T\gsim 170~$MeV the pressure calculated in the SHMC
model  is much higher than that predicted by lattice simulations,
that may give rise to problems in constructing a deconfinement
transition if one wants to match this hadron EoS with a
quark-gluon one at $T\gsim 170~$MeV. The ways out of this problem
have been discussed. The most natural way seems to introduce a
different scaling law for  baryon resonance coupling constants, as
compared to those used for nucleons,  resulting in an increase of
the masses of baryon resonances in high temperature region
compared to those used here. Then we could match our SHMC model
EoS with that for quarks and gluons at quite high temperatures $T>
T_{\rm dec}$. In this case the quark liquid would masquerade as
the hadron one. A similar idea has been recently discussed in
application to hybrid stars in \cite{Alf05}. We demonstrated this
idea by suppressing couplings of all baryons by factor $1/3$,
except nucleons. Then we easily match the lattice pressure up to
$T\sim 240$~MeV ($T_{c\sigma}\sim 330$~MeV for the case of
suppressed couplings by factor $1/3$, except nucleons).

The specific heat in SHMC model generates a sharp peak at $T\sim
190$~MeV simulating a strong crossover behavior. The peak is
smeared out for suppressed couplings. It is interesting that at
$T>T_{c\sigma}$ the squared effective mass of the $\sigma$
excitation becomes negative. This can be interpreted as the
occurrence of the instability with respect to a hot Bose
condensation of $\sigma$ excitations. Similar possibility has been
found in~\cite{V04} in the framework of a different model and it
was named as hot Bose condensation since the condensate, which
stabilizes the system, appears for $T>T_{c}$ rather than for
$T<T_{c}$. Thus there arises a question on a competition between
the hadron hot Bose condensation phase and the deconfined
quark-gluon phase. All this may indicate that the description of
the strongly interacting matter at high temperatures is more
complicated than it was expected. In a more realistic model one
should incorporate both hadron and quark-gluon degrees of freedom
interacting with each other within a strongly correlated mixed
quark-hadron state. Indeed, this idea is as old as twenty
years~\cite{Hatsuda} and recent quenched lattice QCD results show
the existence of resonance structures above $T_c$.

As an example for implementation of the SHMC model  to HIC we
examined the isentropic regime of this EoS which corresponds to
expansion stage of  nuclear system. A new method for extracting
thermodynamic parameters of the freeze-out stage was considered
using a shock-like model of freeze-out \cite{Bug}. This method
takes into account some elements of HIC dynamics and in-medium
hadron modification. Obtained parameters prove to be a little bit
different from those derived by the standard statistical model
procedure for the energy range below the top AGS energy. The
difference is within  error bars for $T_{fr}$ and slightly above
error bars for $\mu_{fr}$ extracted from data by the standard
procedure. For more definite conclusions more precise data are
needed as well as measurements of multistrange particles and
inclusion of them  in the scheme. Moreover, the results depend on
the assumed model of the freeze-out. If we used the prompt
freeze-out concept ~\cite{MSTV90,VS87} we would obtain
significantly larger difference between in-medium and ideal gas
yields.

If we want to apply the SHMC model  for analyzing the particle
ratios in the RHIC energy range, we should extend the used
particle set to higher meson and baryon resonances. Here we don't
do that since their coupling constants are unknown and we won't
generate extra uncertainties. In addition, our aim is to
demonstrate ability of the model in a broad $(T,\mu_B)$ range
rather than to study specific regimes. We are planning to return
to this question in future publication.

As we mentioned, if one moves to still higher temperatures,
quark-gluon degrees of freedom become deciding. We plan to match
our hadron SHMC EoS with quark-gluon EoS developing either a
masquerade or a mixed phase scenario to cover a high $T-\mu_B$
range of the phase diagram. Then, we would like to insert this EoS
into hydrodynamic codes~\cite{TS,IRT06} to study dynamics of HIC.

The SHMC model allows for arbitrary isotopic composition. It is
interesting to apply it to collisions of isospin-asymmetric nuclei
for studying isotopic effects and for describing initial stage of
neutron star formation and cooling process (at temperatures
$T\lsim 50$~MeV and high baryon densities).

The model allows one a further development. It is well known
\cite{KV03,MSTV90} that pions and kaons have strong $\pi NN$, $\pi
N\Delta$, $KN\Lambda$, $KN\Sigma^*$ interactions in the $p$-wave
that we disregarded including only $s$-wave terms. These
interactions can be incorporated into the scheme. For the sake of
simplicity, the gas of quasiparicle excitations was treated as a
non-interacting gas. Feasibly this assumption is unrealistic when
we deal with high temperatures. In the paper body we showed how
one can generalize the model to the Hartree level. This can be
done straightforward. However using RMF models and their
generalizations we should always balance between a realistic and a
practically tractable descriptions. Thus we postpone  with further
generalizations of the SHMC model to future work.

Results are sensitive to the concept of freeze-out and it would be
attractive to probe different freeze-out models.

We used spatially homogeneous solutions. However initial equations
of motion for mean fields are written in coordinate space. Thus in
principle model allows to describe structures of possible mixed
phases that may arise in neutron stars (cf. \cite{MVTC05}) and
feasibly in heavy ion collisions. With spatially inhomogeneous
solutions at hand one could apply the model to study atomic nuclei
and confront it with many  new experimental constraints.

These interesting questions are however beyond the scope of the
present paper and need further investigation.

\vspace*{5mm} {\bf Acknowledgements} \vspace*{5mm}

We are very grateful to  E.E.~Kolomeitsev for numerous
illuminating discussions, valuable remarks and constructive
criticism. We also acknowledge discussions with A.~Andronic,
 B. Friman, Yu.B.~Ivanov, and V.V.~Skokov. This work was supported in part by
the Deutsche Forschungsgemeinschaft (DFG project 436 RUS
113/558/0-3), the Russian Foundation for Basic Research (RFBR
grants 06-02-04001 and 05-02-17695), the Russian President program
``Support of leading scientific schools'' 320.2006.2, and by a
special program of the Ministry of Education and Science of the
Russian Federation (grant RNP.2.1.1.5409).

{\bf{Appendix A.
Energy density of the gas of  boson excitations}}

Here we present expressions for the partial contributions to the energy
density of the gas of  boson excitations, Eq. (\ref{exden}).

The energy density of the $\sigma$ meson excitations on the ground
of the $\sigma$ mean field is
\be
E_{\sigma}^{\rm part}&=&
\intop_0^{\infty}\frac{\rmd p \ p^2}{2\pi^2}\, \om_{\sigma}(p) \
f_{\sigma} (\om_{\sigma}(p)) ,\ee with the dispersion relation \be
\om_{\sigma}(p)= \sqrt{(m_\sigma^{\rm part *})^2\, +p^2},
\ee
 and
the effective mass (\ref{spa}):
\be
\label{sigm-mass} (m_{\sigma}^{\rm part
*})^2 &=&\frac{C_{\sigma}^2 \  m^2_{\sigma} \
\Phi^2_{\sigma}(f)}{m_N^4 \ \eta_{\sigma}} \
\frac{\prt^2 E_{\rm MF}[f,\omega_0 (f),R_0 (f), T]}{\prt f^2}\nonumber\\
&&\times \left[
1-f\frac{\Phi^{\prime}_{\sigma}(f)}{\Phi_{\sigma}(f)}
+f\frac{\eta^{\prime}_{\sigma}(f)}{2\eta_\sigma}\right]^{-2}~,
\ee
where $E_{\rm MF}[f,\omega_0 (f),R_0 (f), T]$ is given by
Eqs.~(\ref{Efun}), (\ref{extreme}), (\ref{extremef}). All
derivatives are taken here over variable $f$ with fixed particle
occupation.

The energy density of the vector  $\om$-meson excitations on the
ground of the $\om$ mean field is
\be
E_\om^{\rm part}&=& (2s_\om
+1)\intop_0^{\infty}\frac{\rmd p \ p^2}{2\pi^2}\, \om_{\om}(p) \
f_{\om} (\om_{\om}(p))~,
\ee
with the $\om$ spin $s_\om =1$ and

\be
\label{om-mass} \om_{\om}(p)=\sqrt{(m_\om^{\rm part *})^2\, +p^2},
\quad m_\om^{\rm part
  *} =m_\om |\Phi_\om (f)|~,
  \ee
cf. Eq. (\ref{omr}). Similarly the energy density of the vector
and iso-vector $\rho$-meson excitations on the ground of the
$\rho$ mean field is

\be &&E_\rho^{\rm part}=E_{\rho^{+}}+ E_{\rho^{0}}+E_{\rho^{-}}=(2s_\rho +1)\\
&&\times\intop_0^{\infty}\frac{\rmd p \ p^2}{2\pi^2}\,
[\om_{\rho^+}(p) \ f_{\rho} (\om_{\rho^+}(p))+\om_{\rho^0}(p) \
f_{\rho} (\om_{\rho^0}(p))+\om_{\rho^-}(p) \ f_{\rho}
(\om_{\rho^-}(p))] \nonumber
 \ee
with the $\rho$ spin $s_\rho =1$ and

\be\label{rho-mass}
\om_{\rho^{\pm}}(p)&=&\mp V\pm g_\rho \ \chi^{\prime}_\rho \
R_0 +\sqrt{(m_\rho^{\rm part *})^2\, +p^2},\\
\om_{\rho^0}(p)&=&\sqrt{(m_\rho^{\rm part *})^2\, +p^2}, \quad
m_\rho^{\rm part *} =m_\rho |\Phi_\rho (f)|\,.
\ee

Conditions $\partial_{\mu}\om^{\mu}=0$,
$\partial_{\mu}\vec{\rho}^{\mu}=0$ are  fulfilled.

The energy density of the pion gas is
\be
&&E_{\pi}^{\rm part}=
E_{\pi^{+}} +E_{\pi^{0}} +E_{\pi^{-}}= \intop_0^{\infty}\frac{\rmd
p \ p^2}{2\pi^2}\,\\
&&\times [ \om_{\pi^+}(p) \ f_{\pi^+}
(\om_{\pi^+}(p))+\om_{\pi^0}(p) \ f_{\pi^0} (\om_{\pi^0}(p))+
\om_{\pi^-}(p) \ f_{\pi^-} (\om_{\pi^-}(p))]~, \nonumber
\ee
where
for charged and neutral pions we have, respectively
\be
\om_{\pi^{\pm}}(p)= \mp V\pm g_{\om \pi}^* \ \om_0 \pm
g_{\rho\pi}^* \ R_0 + \sqrt{m_{\pi}^{* 2}\, +p^2} ~, \ee \be
\om_{\pi^0}(p)=\sqrt{(m_{\pi}^{*})^2\, +p^2},\quad
m_{\pi}^{*}=m_{\pi}-g_{\sigma \pi}^*\sigma
\ee
 with $ g_{\om
\pi}^* =0$ due to absence of the $\om\rightarrow 2\pi$ decay.

The energy density of the kaon gas is

\be
&&E_{{K}}^{\rm part}= E_{K^+}+ E_{K^0}+E_{K^-}+E_{\bar{K}^0}\\
&&=\intop_0^{\infty}\frac{\rmd p \ p^2}{2\pi^2}\, [
\om_{K^+}(p) \ f_{K^+} (\om_{K^+}(p))+\om_{K^0}(p) \ f_{K^0} (\om_{K^0}(p))]
\nonumber\\
&&+\intop_0^{\infty}\frac{\rmd p p^2}{2\pi^2}\, [ \om_{K^-}(p) \
f_{K^-} (\om_{K^-}(p))+\om_{\bar{K}^0}(p) \ f_{\bar{K}^0}
(\om_{\bar{K}^0}(p))] ~, \ee where for charged kaons \be
\label{EK+} \om_{K^{\pm}}(p)&=& \mp V\pm g_{\om K}^* \ \om_0 \pm
g_{\rho K}^* \
R_0 + \sqrt{m_{K}^{* 2}\, +p^2} ~, \\ 
m_{K}^{*}&=&m_{K}-g_{\sigma K}^*\sigma \ee

and similarly for neutral kaons \be \label{EK0}
\om_{{K}^0/{\bar{K}^0} }(p)=\pm g_{\om K}^* \ \om_0 \pm g_{\rho
K}^* \ R_0 + \sqrt{m_{K}^{* 2}\, +p^2}~, \ee

The $\eta$ contribution to the energy density, cf. \cite{ZPLN}, is
given by
\be
E_{\eta}^{\rm part}&=& \intop_0^{\infty}\frac{\rmd p
\ p^2}{2\pi^2}\, \om_{\eta}(p) \ f_{\eta} (\om_{\eta}(p))~, \quad
\omega_{\eta}=\sqrt{m_{\eta}^{* 2}+{p}^2}~,
\ee
where
\be
\label{etam} m_{\eta}^{* 2}=\left( m_{\eta}^{2}-\sum_{b\in\{b\}}
\frac{\Sigma_{\eta b}}{f^2_{\pi}} <\bar\Psi_b \,\Psi_b> \right) /
\left( 1+\sum_{b\in\{b\}}\frac{\kappa_{\eta
b}}{f^2_{\pi}}<\bar\Psi_b \,\Psi_b>\right)
\ee
and the total
baryon scalar density is $\sum_{b\in\{b\}}<\bar\Psi_b
\,\Psi_b>=\sum_{b\in\{b\}}n_b^{\rm sc}$.

The Bose distributions of
 excitations are
\be
\label{Tbos} f_{\rm i} &=&\frac{1}{\mbox{exp}[(\sqrt{m_{\rm
i}^{*2}+p^2}- \mu_{\rm i}^*)/T]-1},\\ \label{Tmubos} \mu_{\rm i}^*
&=&\mu_{\rm i} +Q_{\rm i} (\mu_{\rm
  ch}+V)-Q^{\rm vec}_{\rm i} \ g_{\om {\rm i}}^* \
\om_0 -Q^{\rm vec}_{\rm i} g_{\rho {\rm i}}^* R_0 ,\\ i\in\{{\rm
b.ex}\}&=& \sigma ,\om,\rho^+ ,\rho^0 , \rho^-  ; \pi^{+},\pi^0
,\pi^- ; K^+ ,K^0 , K^- ,\bar{K}^0 ; \eta ;\nonumber\\
 && K^{*+} ,K^{*0} , K^{*-} ,
\bar{K}^{*0}; \eta^{'}; \varphi ~.\nonumber \ee Here $\mu_i
=\mu_{\rm str}$  for strange particles $K$ and $K^{*}$ and $\mu_i
=-\mu_{\rm str}$ for their antiparticles. $Q_{\rm i}$ is the boson
electric charge in proton charge units, and we again did a gauge
shift of the $V$ variable, $Q^{\rm vec}_{\rm i} =+1$ for particle,
$Q^{\rm vec}_{\rm i} =-1$ for antiparticle and $Q^{\rm vec}_{\rm
i} =0$ for the neutral particles (with all zero charges including
strangeness). We take $g_{\om \pi}=0$, $g_{\om K^{*}}= g_{\rho
K^{*}}=0$, $g_{\om i}=0$ for $i=\om$ and $g_{\rho i}=0$ for
$i=\rho$. For simplicity $m^{*}_{K^{*}}=m_{K^{*}}$,
$m^{*}_{\eta^{'}}=m_{\eta^{'}}$, $m^{*}_{\varphi}=m_{\varphi}$ are
assumed due to absence of corresponding experimental data.

Note that the inclusion of the $p$-wave pion and kaon terms can be
easily done. For that one  needs to replace $\om_{\pi}(p)$ and
$\om_{K}(p)$ to more complicated expressions which can be found in
Refs. \cite{KV03,MSTV90,KVK96}.
 However, we disregard such an interaction
since its inclusion is  beyond the scope of our RMF based
scheme.

{\bf{Appendix B. Condensation of  excitations}}

 There are two possible types of
condensations of excitations. One type is the Bose-Einstein
condensation which might occur for particles with a conserved
charge described by a finite value of the corresponding chemical
potential. The number of condensed particles is determined by the
value of the conserved charge (electric and/or strangeness in our
case) even if one ignores self-interaction of bosons. Another type
of condensation might occur for neutral particles. For them
chemical potential is zero. The stability in this case is achieved
by the interaction of the given boson species with other particle
species and by the self-interaction.

{\bf{ Bose-Einstein condensation.}} Let us consider charged bosons
(a charge means either electric charge or strangeness). The
density of the gas of the Bose excitations of the given species
$i$ is determined by
 the integral
\be
\label{bosun}
n_i =g_i \intop_0^{\infty}\frac{\rmd p \
p^2}{2\pi^2}\,
f_i(p)~,
\ee
where $g_i$ is the degeneracy factor and $f_i$ is
defined by Eq.~(\ref{Tbos}). For $T>T_{ci}$ we have $\mu_i^*
(T)<m_i^{*}(T)$ and there is no Bose-Einstein condensation. The
critical point of the Bose-Einstein condensation is determined by
the condition $\mu_i^* (T_{ci} )=m_i^{*}(T_{ci})$. For  $T<T_{ci}$
the  integral (\ref{bosun}) diverges. Thus for $T<T_{ci}$ the
excitation number density $n_i$ acquires a condensate term
\be
\label{bose} n_i =g_i \intop_0^{\infty} \
 \frac{\rm d p \
 p^2}{2\pi^2}\, f_i(p)+n_{0i}~.
\ee
Assume that the  non-linear self-interaction term in the
Lagrangian density is omitted. Then the relation $n_{0i}=2m_i^*
(T)|\phi^{\rm cl}|^2$ connects the condensate density   and the
mean field $\phi^{\rm cl}$ (for simplicity we assumed the presence
of a single condensate field). Then the energy density acquires
extra term \be \delta E =\sum_{i\in\{ex\}} n_{0i} (n_i ,T)m_i^* .
\ee
 The condensate does not contribute
to the pressure. Contribution to the entropy of the Bose-Einstein
condensate is also zero. Since $-\mu_i^* n_i  =-m_i^* n_i$ for
$T<T_{ci}$,  the thermodynamic consistency relation
holds.

Generalization to the case, when  the self-interaction term in the
Lagrangian density is included, can be done as follows. The
relation between $\mu_i^*$ and the mean field $\phi^{\rm cl}$ is
found from the equation of motion for the mean field. As we have
mentioned we should present the given boson field as $\phi
=\phi^{\rm cl}+\phi^{\prime}$. In the gas approximation for
excitations we should keep only quadratic terms ($\propto
|\phi^{\prime}|^2 $) in the Lagrangian density. Consider again a
single (quasi)Goldstone excitation. By averaging $\mathcal{L}^{\rm
int}_{\rm G} (\phi^{\prime})$ over the equilibrium state we find
\be
\label{mespot1} &&\delta U = <\mathcal{L}^{\rm int}_{\rm G}
(\phi^{\prime})> \simeq \lambda |\phi^{\rm cl}|^4 /2 +2\lambda
|\phi^{\rm cl}|^2 <|\phi^{\prime}|^2 > , \\
&& <|\phi^{\prime}|^2 >=n^{\rm sc}_{\phi}~.
\ee
where $n^{\rm
sc}_{\phi}$ is the boson scalar density (sum of tadpole diagrams)
for the given species.

Minimization of the total energy results in equations of motion
for the classical field and excitations. For classical field we
obtain:
\be
\label{clmes} |\phi^{\rm cl}|^2 =(\xi /\lambda )\theta
(\xi), \ee where \be (\om_c +\mu^* )^2 -m^{*2}-2\lambda n^{\rm
sc}_{\phi}\equiv \xi ,
\ee
and $\om_c $ is the critical frequency
given by $\om_c +\mu^*_c =m^*_c$. If we replace $\om_c +\mu^*_c$
by $\mu_{c}^* (T_{c})$ and  put $\lambda =0$ we  recover the
Bose-Einstein condensation result discussed above.

The spectrum of particles now becomes
\be
(\om +\mu^* )^2 =m^{*2}
+\vec{p}^{\,2} +2\lambda |\phi^{\rm cl}|^2 ~.
\ee
Thus as compared
to calculations done in absence of the condensate, we should
replace $\om = \sqrt{m^{*2} +\vec{p}^{\,2}} -\mu^*$ by $\om =
\sqrt{m^{*2} +2\lambda |\phi^{\rm cl}|^2 +\vec{p}^{\,2}} -\mu^*$
in equations for particle excitation energies and  add the terms
$\lambda |\phi^{\rm cl}|^4 /2$ and $-\lambda |\phi^{\rm cl}|^4 /2$
to the energy density and pressure, respectively. For charged
pions the chiral symmetry consideration provides the relation
$\lambda =m_{\pi}^2 /f_{\pi}^2$.

{\bf{Condensation of neutral bosons.}} For neutral bosons, if
strangeness and electric charge  are zero, condensations may also
appear. In this case $\mu_i^* =0$. The critical point is
determined by the condition $m^* (T_{ci})=0$. The stability is
achieved only due to the non-zero meson-meson self-interaction
with the positive coupling constant $\lambda_i$ for $\phi^4$ term.
For $\eta^{'}$ and $\varphi$ we use free dispersion relations,
thereby they never condense in the framework of our consideration.

In spite of the fact that condensate of the $\sigma$-excitations
appears for $T>T_{c\sigma}$ rather than for $T<T_{c\sigma}$, it
can be considered in the similar way as the cases discussed above.


\end{document}